%% file: ms_M_stars.tex
\documentclass{emulateapj}
\usepackage{graphicx}
\usepackage{amsmath}

\usepackage{color}

\newcommand{\appropto}{\mathrel{\vcenter{
  \offinterlineskip\halign{\hfil$##$\cr
    \propto\cr\noalign{\kern2pt}\sim\cr\noalign{\kern-2pt}}}}}

\def\kms{kms$^{-1}$\,}

\def\kms{\ifmmode{\rm km\thinspace s^{-1}}\else km\thinspace s$^{-1}$\fi}

\shortauthors{Rappaport et al.~2013}
\shorttitle{Multiple M-Star Systems}

\begin{document}

%
\def\ltsima{$\; \buildrel < \over \sim \;$}
\def\lsim{\lower.5ex\hbox{\ltsima}}
\def\gtsima{$\; \buildrel > \over \sim \;$}
\def\gsim{\lower.5ex\hbox{\gtsima}}
%

\bibliographystyle{apj}

\title{
M-Dwarf Fast Rotators and the Detection of Relatively Young Multiple M-Star Systems
}

\author{
S.~Rappaport\altaffilmark{1},
J.~Swift\altaffilmark{2},
A.~Levine\altaffilmark{3},
M.~Joss\altaffilmark{1},
R.~Sanchis-Ojeda\altaffilmark{1},
T.~Barclay\altaffilmark{4}, \\
M.~Still\altaffilmark{4},
G.~Handler\altaffilmark{5}, 
K.~Ol\'ah\altaffilmark{6},
P.\,S.~Muirhead\altaffilmark{7,8}
D. Huber\altaffilmark{9}, and
K. Vida\altaffilmark{6}
}

\altaffiltext{1}{Department of Physics, and Kavli Institute for
  Astrophysics and Space Research, Massachusetts Institute of
  Technology, Cambridge, MA 02139, USA; sar@mit.edu, mattjoss@mit.edu, rsanchis86@gmail.com}
  
\altaffiltext{2}{Department of Astronomy and Department of Planetary Science, California Institute of Technology, MC 249-17, Pasadena, CA 91125, USA; jswift@astro.caltech.edu}

\altaffiltext{3}{37-575 M.I.T.\ Kavli Institute for Astrophysics and Space Research, 70 Vassar St., Cambridge, MA, 02139; aml@space.mit.edu}

\altaffiltext{4}{BAER Institute/NASA Ames Research Center, M/S 244-30, Moffett Field, Mountain View, California 94035; thomas.barclay@nasa.gov, martin.d.still@nasa.gov}

\altaffiltext{5}{Nicolaus Copernicus Astronomical Center, Bartycka 18, PL 00-716 Warsaw, Poland; gerald@camk.edu.pl}

\altaffiltext{6}{Konkoly Observatory of the Hungarian Academy of Sciences, PO Box 67, H-1525 Budapest, Hungary;  olah@konkoly.hu; vidakris@konkoly.hu}

\altaffiltext{7}{Department of Astronomy, Boston University, Boston, MA 02215; philipm@bu.edu}
\altaffiltext{8}{Hubble Postdoctoral Fellow}
\altaffiltext{9}{SETI Institute/NASA Ames Research Center, MS 244-30, Moffett Field, CA 94035, USA; daniel.huber@nasa.gov}

\begin{abstract}

We have searched the {\em Kepler} light curves of $\sim 3900$ M-star
targets for evidence of periodicities that indicate, by means of the
effects of starspots, rapid stellar rotation.  Several analysis
techniques, including Fourier transforms, inspection of folded light
curves, `sonograms', and phase tracking of individual modulation
cycles, were applied in order to distinguish the periodicities due to
rapid rotation from those due to stellar pulsations, eclipsing
binaries, or transiting planets.  We find 178 {\em Kepler} M-star targets with rotation
periods, $P_{\rm rot}$, of $<$ 2 days, and 110 with $P_{\rm rot}<$ 1
day.  Some 30 of the 178 systems exhibit two or more
independent short periods within the same {\em Kepler} photometric
aperture, while several have three or more short periods.  Adaptive
optics imaging and modeling of the {\em Kepler} pixel response
function for a subset of our sample support the conclusion that the
targets with multiple periods are highly likely to be relatively young
physical binary, triple, and even quadruple M star systems.  We
explore in detail the one object with four incommensurate periods all
less than 1.2 days, and show that two of the periods arise from one of
a close pair of stars, while the other two arise from the second star,
which itself is probably a visual binary. If most of these M-star systems 
with multiple periods turn out to be bound M stars, this could prove 
a valuable way of discovering young hierarchical M-star systems; the same
approach may also be applicable to G and K stars. The $\sim$5\% 
occurrence rate of rapid rotation among the $\sim 3900$ M star targets 
is consistent with spin evolution models that include an initial 
contraction phase followed by magnetic braking, wherein a typical M
star can spend several hundred Myr before spinning down to periods
longer than 2 days.

\end{abstract}


\section{Introduction}

M dwarfs---main-sequence stars with masses $M_\star \lesssim 0.6
\,M_\odot$---dominate the Galactic stellar population in number and
mass (Chabrier 2003).  This preponderance is reflected in the local 
neighborhood where $\sim$$75$\% of the nearest stars fall within this spectral
class (Henry et al. 1994, 2006; Reid et al. 2002; Cruz \& Reid 2002), and this 
fraction will continue to grow as more nearby red dwarfs are revealed. Despite 
their ubiquity, these small stars are difficult to characterize due in part to their 
low luminosities and complex atmospheres, and models are unable to 
satisfactorily account for their radii and temperatures at given
masses (Torres et al.~2010; Boyajian et al.~2012; Birkby et al.~2013). 
The motivation to understand the fundamental properties of
M dwarfs, including their multiplicity, has been amplified by recent
results indicating that they frequently host low-mass planets
(Swift et al.~2013; Dressing \& Charbonneau~2013; 
Morton \& Swift~2013).

Low-mass stars lose angular momentum through rotational magnetic braking 
(Mestel 1968). However, during their pre-main sequence lifetime their rotation 
rates increase due to gravitational contraction. This pre-main sequence phase 
can last from $\sim$100 Myr for stars of mass 0.5 $M_\odot$ to several hundred Myr for the 
lowest mass stars (see, e.g., Baraffe et al.~2002). After hydrogen burning commences 
in their cores and gravitational contraction ceases these stars begin to gradually spin 
down. Consequently, their magnetic field strengths and activity levels tend to decline 
(Skumanich 1972).

M dwarfs are thought to rotate differentially, but less so than solar
type stars. Recent measurements based on {\em Kepler} data show that
differential rotation rates for M dwarfs are typically less than 0.1
rad\,d$^{-1}$ (Reinhold et al.~2013). Observations of late type stars
show that their differential rotation rates generally 
depend strongly on effective
temperature and weakly on rotation period
(Barnes et al.~2005; Reiners 2006), a behavior that is
also seen in computer models (K\"{u}ker \& R\"{u}diger 2011).

The {\em Kepler} Space Mission (Borucki et al.~2010; Koch et al.~2010)
monitored over 150,000 stars nearly continuously for about four years,
primarily to search for exoplanets (Borucki et al. 2011; Batalha et
al. 2013; Burke et al. 2013). {\em Kepler}'s unprecedented photometric
precision has also led to extensive ancillary
investigations\footnote{\url{keplerscience.arc.nasa.gov/PublicationsAstrophysics.shtml}}.

Though the vast majority of {\em Kepler} target stars are Sun-like
($0.8 \lesssim M_\star \lesssim 1.2 \, M_\odot$), several thousand M
dwarfs have been monitored as well over the course of the primary
mission. The initial estimation of the important characteristics of
the stars in the {\em Kepler} field was done using ground-based
multicolor photometry as part of the construction of the {\em Kepler}
Input Catalog (KIC).  This estimation process was optimized for
sun-like stars; the characteristics of the M dwarfs in the KIC were not 
determined as accurately and should be adopted only with caution (Brown 
et al.~2011). There have been several efforts to revise the stellar
parameters for this M-star sample (e.g. Muirhead et
al.~2012a; Mann et al.~2012, 2013). In particular, Dressing \&
Charbonneau (2013; hereafter ``DC13'') have tabulated and calibrated
the properties of 3897 cool and high surface gravity {\em Kepler}
target stars. The vast majority of these have masses in the range $0.3
< M_\star < 0.6 \, M_\odot$ and the bulk of the effective temperatures
fall in the range $3300 < T_{\rm eff} < 4000$ K.  The numbers of stars
show a bias toward the higher mass M dwarfs due to the magnitude
limited nature of the {\em Kepler} targets (Batalha et al.~2010). Some
90\% of the DC13 collection of M stars are at distances in the range
15-400 pc (median distance of 210 pc), and have absolute visual 
magnitudes in the range +8 to +12.

\begin{figure}
\begin{center}
\includegraphics[width=0.98 \columnwidth]{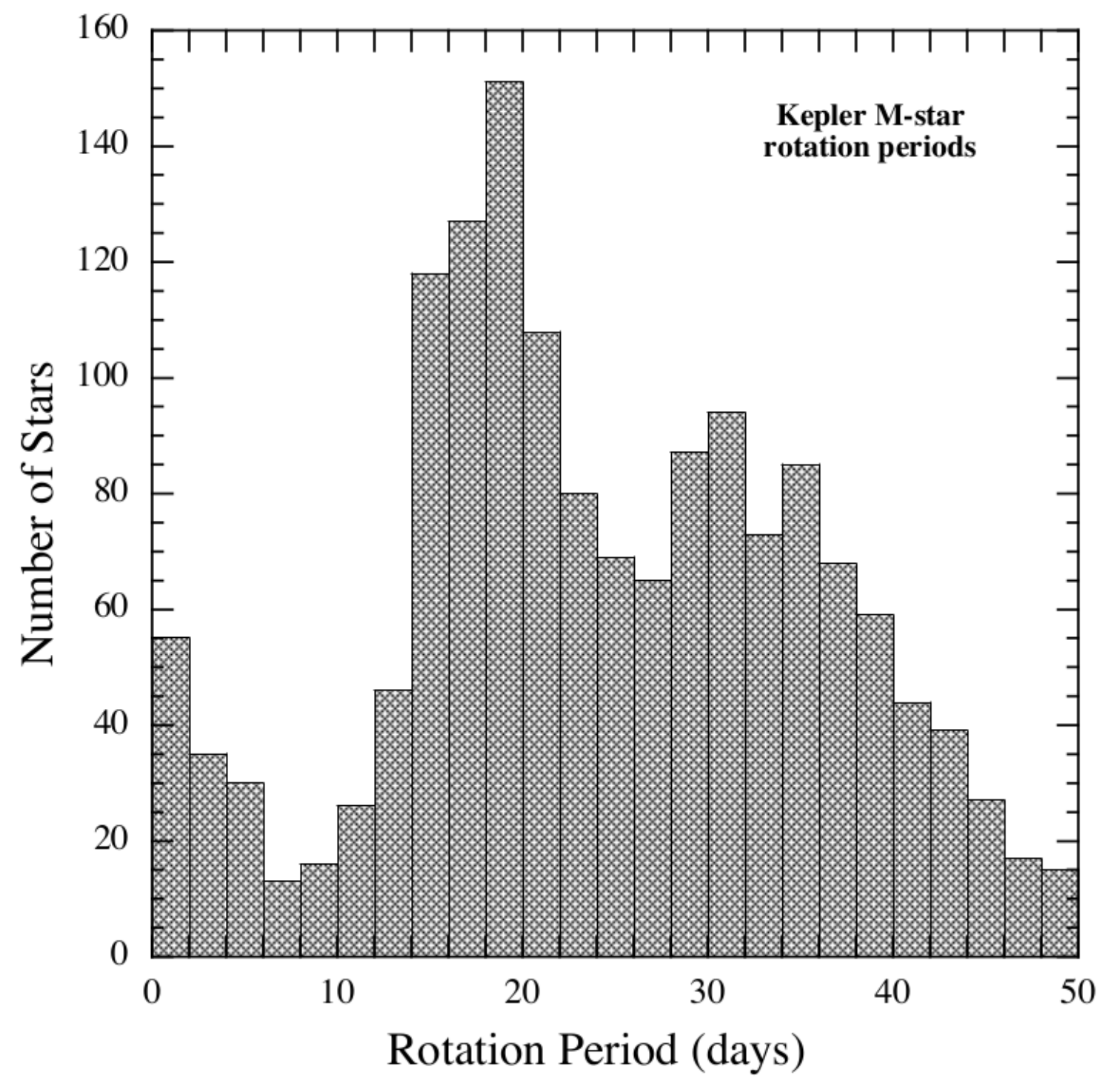} \vglue0.2cm  
\includegraphics[width=0.95 \columnwidth]{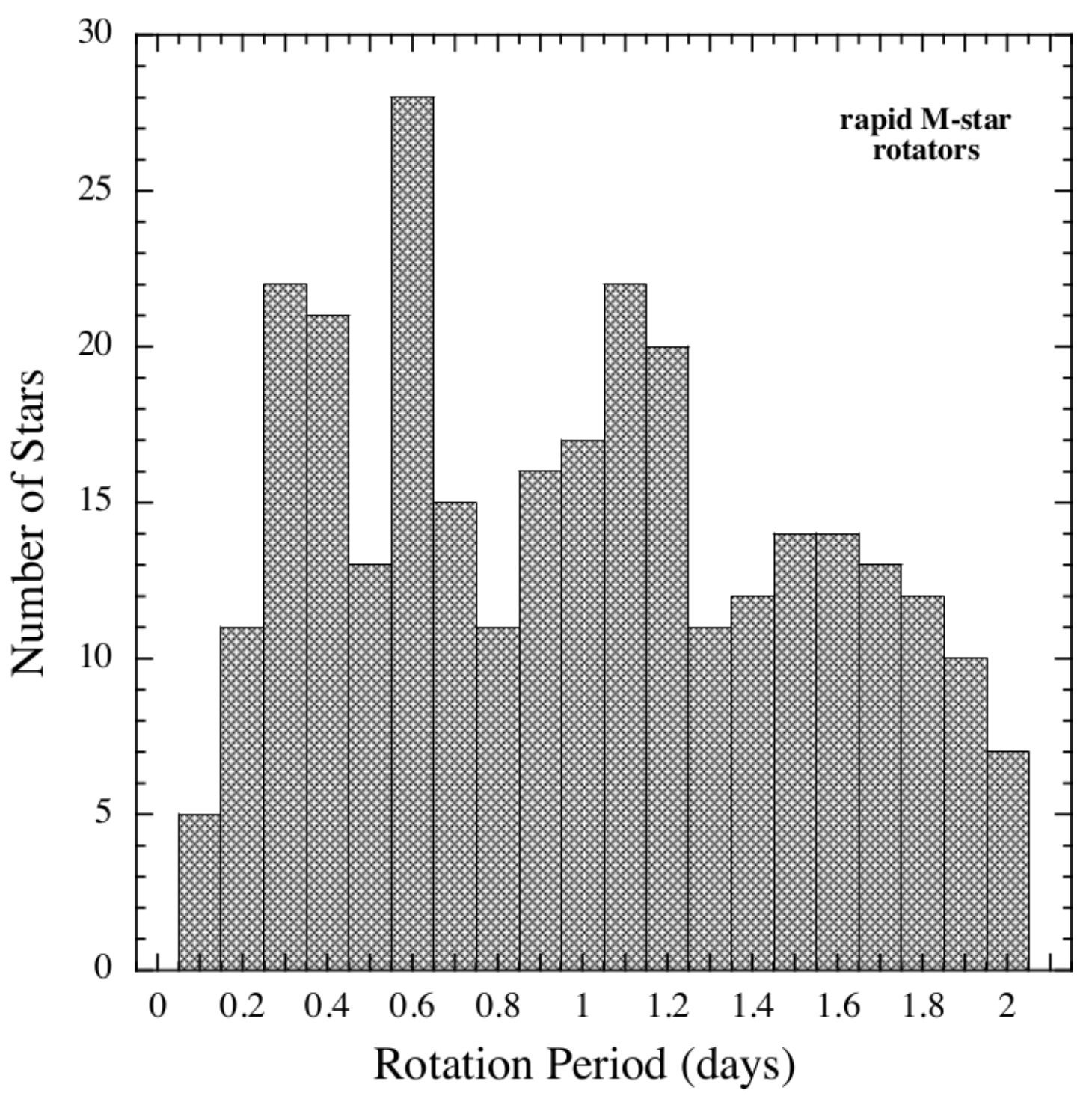}			
\caption{Top: Rotation period distribution of M stars in the {\em Kepler} data set; the periods were taken from McQuillan et al.~(2013), and are based on 10 months of data.  Note the distinct group of relatively rapid rotators with periods shorter than $\sim$5 days.  Bottom: Distribution of very short rotation periods found in {\em Kepler} M-star targets via our Fourier transform search of all 16 quarters of data.  In the case of multiple periods in a single {\em Kepler} target, we include {\em all} the periods in this histogram.}
\label{fig:periods}
\end{center}
\end{figure}

McQuillan, Aigrain, \& Mazeh (2013) performed both autocorrelation
function (`ACF') and Fourier transform analyses of the {\em Kepler}
photometric data for the M dwarf targets in order to identify
variations of the stellar fluxes due to starspots and to measure the
corresponding periods.  The period distribution that they derived is
shown in the top panel of Fig.~\ref{fig:periods}. The distribution
includes periods all the way down to less than a day, has local maxima
near 19 days and 33 days, and tails off dramatically toward periods
above 40 days.  There is also what appears to be a distinct group of M
stars with periods less than $\sim$6 days, and the distribution is
rising toward the shortest periods.  Some 55 of these M stars have
rotation periods of $\lesssim 2$ days.

In this work we focus on the rapidly rotating M stars, in particular
those with $P_{\rm rot} < 1$ day.  In Section~\ref{sec:search} we
describe our search through the {\em Kepler} photometric data base for
rapidly rotating M stars using a Fourier transform algorithm.
In Section~\ref{sec:pulsations} 
the possibility that some of the periodicities we observe are due to
stellar pulsations is discussed and largely discounted.  We show selected
examples of `sonograms' for several of our candidate
rotators in Section~\ref{sec:TFA}, and demonstrate that various
frequencies and their harmonics appear to vary independently,
thereby arguing for the rotating spotted star hypothesis.  In 
Section~\ref{sec:phases} we present an analysis where we  
track the phases of individual starspot rotation cycles and thereby 
demonstrate that these are not stellar pulsations.  In
Section~\ref{sec:multiple} we discuss the {\em Kepler} M-star targets 
wherein we have found two or more distinct rotation periods.
This includes some 30 with at least two short periods, several with 
three periods, and one with four distinct periods (all shorter than 1.2 
days).  We argue in Section~\ref{sec:hierarchical} that these are 
actually binary, triple, and quadruply bound M-star systems.  In 
Section~\ref{sec:images}, we present evidence, based on UKIRT J-band
and Keck adaptive optics images, that these are indeed binary and/or 
hierarchical systems.  In Section~\ref{sec:4660} we apply a 
point-spread-function analysis to the {\em Kepler} pixel-level data 
for KIC 4660255 which exhibits four short periodicities, and show that 
two of them arise in each of two stellar images separated by about 
4$''$.  We summarize our results and draw some final conclusions in 
Section~\ref{sec:conclusions}.

\begin{figure}
\begin{center}
\includegraphics[width=0.97 \columnwidth]{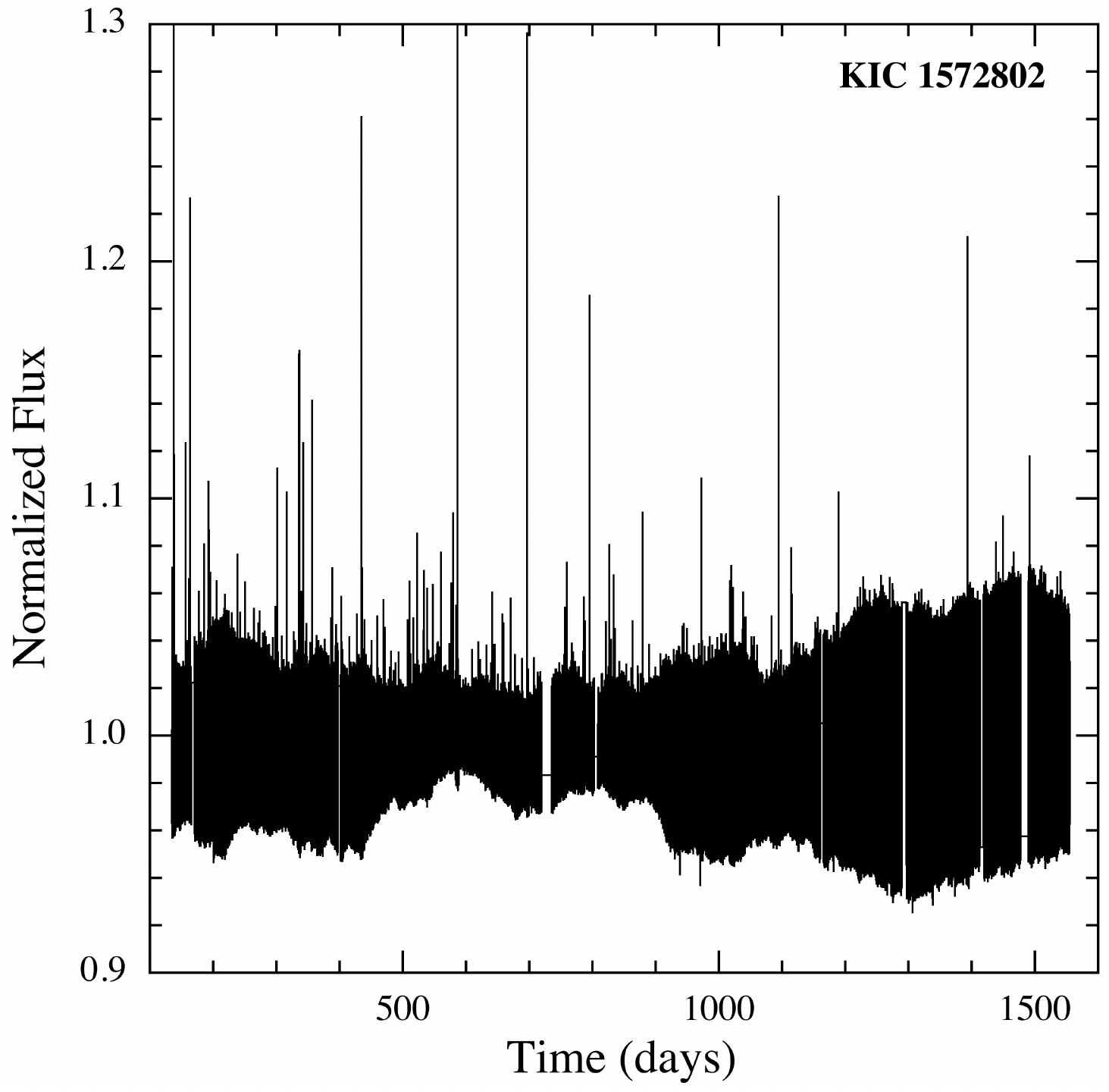} \vglue0.2cm  
\includegraphics[width=0.97 \columnwidth]{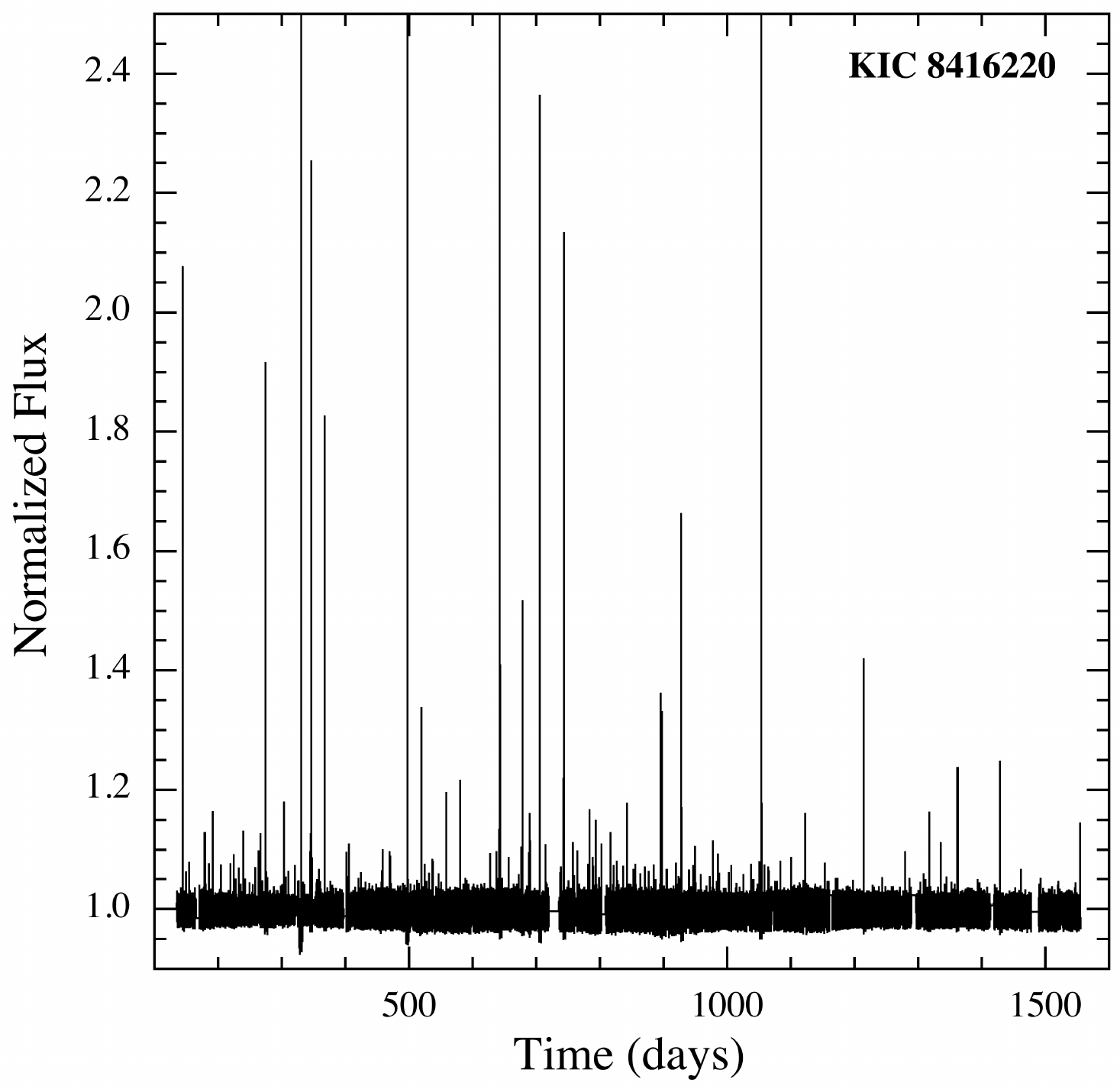}
\caption{Illustrative {\em Kepler} light curves of M-stars.  Top: KIC 1572802 which exhibits flares as large as $\sim$30\% of the mean stellar flux, and one periodicity of 0.37 days.  Bottom: KIC 8416220, which has two periods of 0.72 and 0.57 days, and exhibits flares that increase the flux by a whole magnitude. }
\label{fig:LC}
\end{center}
\end{figure}

\section{Search for Rapidly Rotating M Stars}
\label{sec:search}

Because the Fourier transform (`FT') is an efficient tool for searching
for periodic signals with high-duty cycle and smoothly varying
profiles (i.e., where the first few harmonics dominate), it was our
choice for discovering spotted stars rotating with short periods.  While the
transitory nature of starspots, which can induce possible
erratic changes in the modulation phase, and surface differential
rotation can broaden the peaks in a Fourier power spectrum,
we did not find it necessary, nor particularly useful, to
employ an ACF analysis.  As it turns out, we have discovered a
substantial number of {\em Kepler} M-star targets that exhibit more
than one independent period---sometimes close to or near multiples
of one another---and the FT is substantially more straightforward
than an ACF to use in identifying multiple periods.

\begin{figure}
\begin{center}
\includegraphics[width=0.97 \columnwidth]{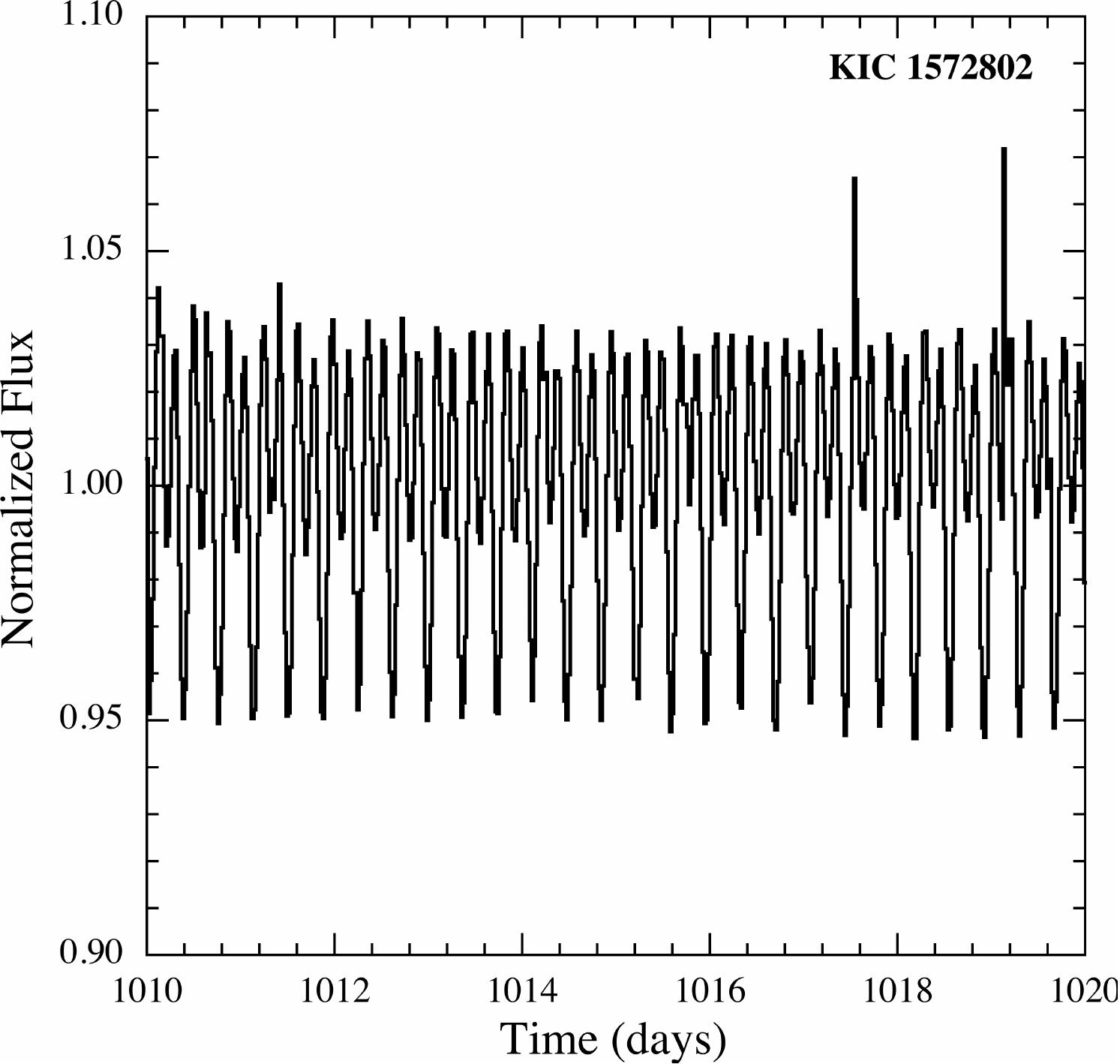} \vglue0.2cm 
\includegraphics[width=0.97 \columnwidth]{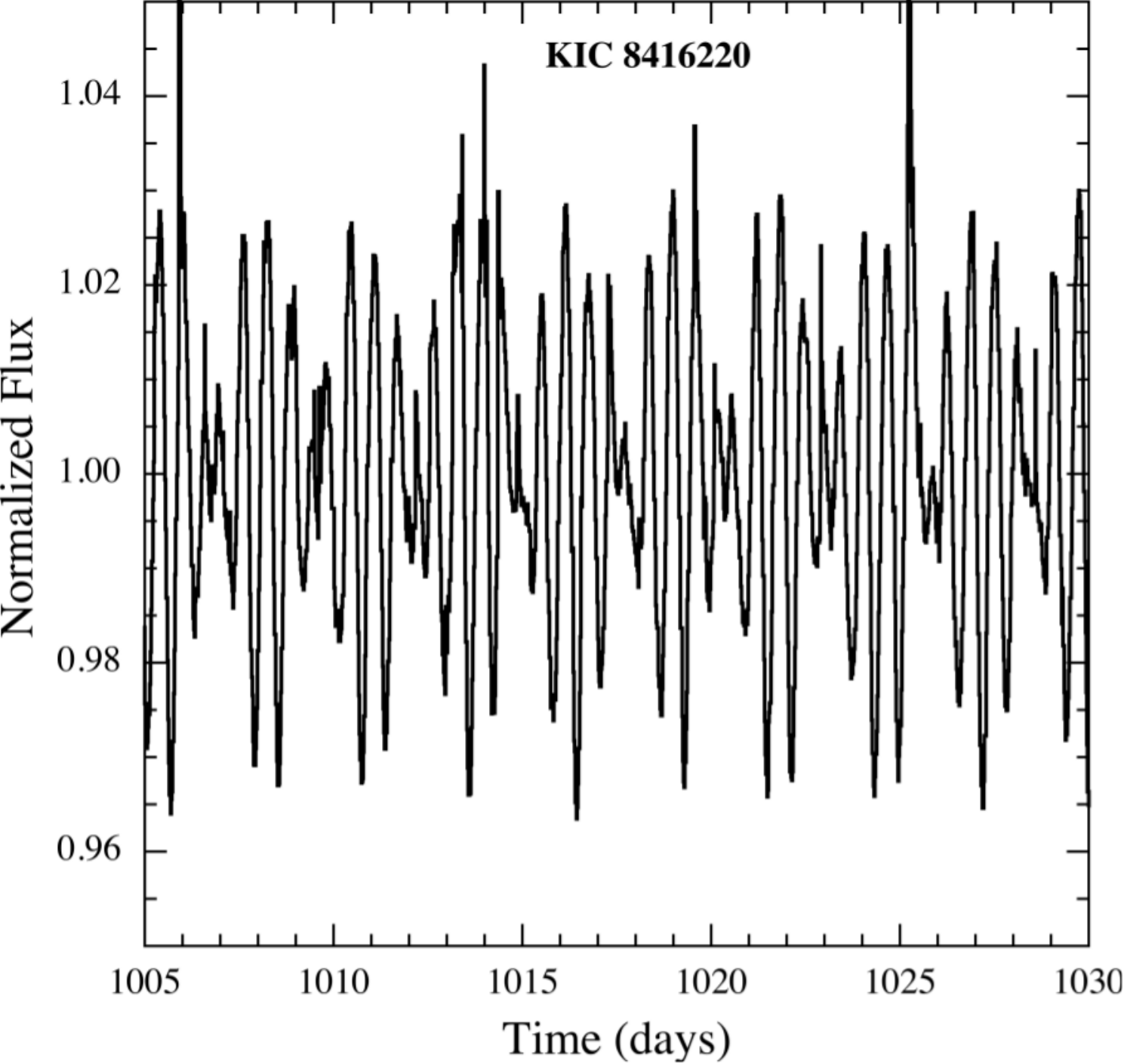} 
\caption{Short segments of the light curves for KIC 1572802 and KIC 8416220 shown in Fig.~\ref{fig:LC}.  Top: The structure in the light curve of KIC 1572802 results from a single periodicity, but with higher harmonic structure (see Fig.~\ref{fig:FTs}). Bottom: The structure in the light curve of KIC 8416220 results from the beat of two periodicities that differ by $\sim$23\% (see Fig.~\ref{fig:FTs}).}
\label{fig:LCz}
\end{center}
\end{figure}

The FT approach utilized in this work is very similar in nature to the 
one used by Sanchis-Ojeda et al.~(2013) in their search for short-period 
planets, and is based on the same selection criteria.  
In brief, the available {\em Kepler} PDCSAP\!\_FLUX data 
(corrected with PDCMAP; Stumpe et al.~2012; Smith et al.~2012) were 
normalized quarter-by-quarter with the quarterly median values and then stitched together into a 
single data file, with all data gaps filled with the mean flux.  After the FTs 
were computed, {\em amplitude} spectra were produced with units of parts per
million (ppm) variation with respect to a flux of unity.  We refer to these Fourier 
transforms below simply as ``the FT''.   For the purpose of searching for statistically 
significant peaks, we further normalized the FT by dividing by a smoothed version of
the FT.  The smoothing was accomplished by convolution with a boxcar 
function that is 100 frequency bins in length.  This procedure has the 
effect of normalizing the raw FT to its local (100-bin) mean.  All targets 
whose normalized FT revealed at least one frequency peak 
exceeding the local mean by a factor $>4$ with at least one harmonic or 
subharmonic which exceeded its local mean by a factor of 3, were 
considered to be worthy of further investigation.
We note that the analysis outlined here of the 3897 DC13 M-star {\em Kepler} targets requires only about one hour
of cpu time on a standard laptop machine.

\begin{figure*}
\begin{center}
\includegraphics[width=0.496 \textwidth]{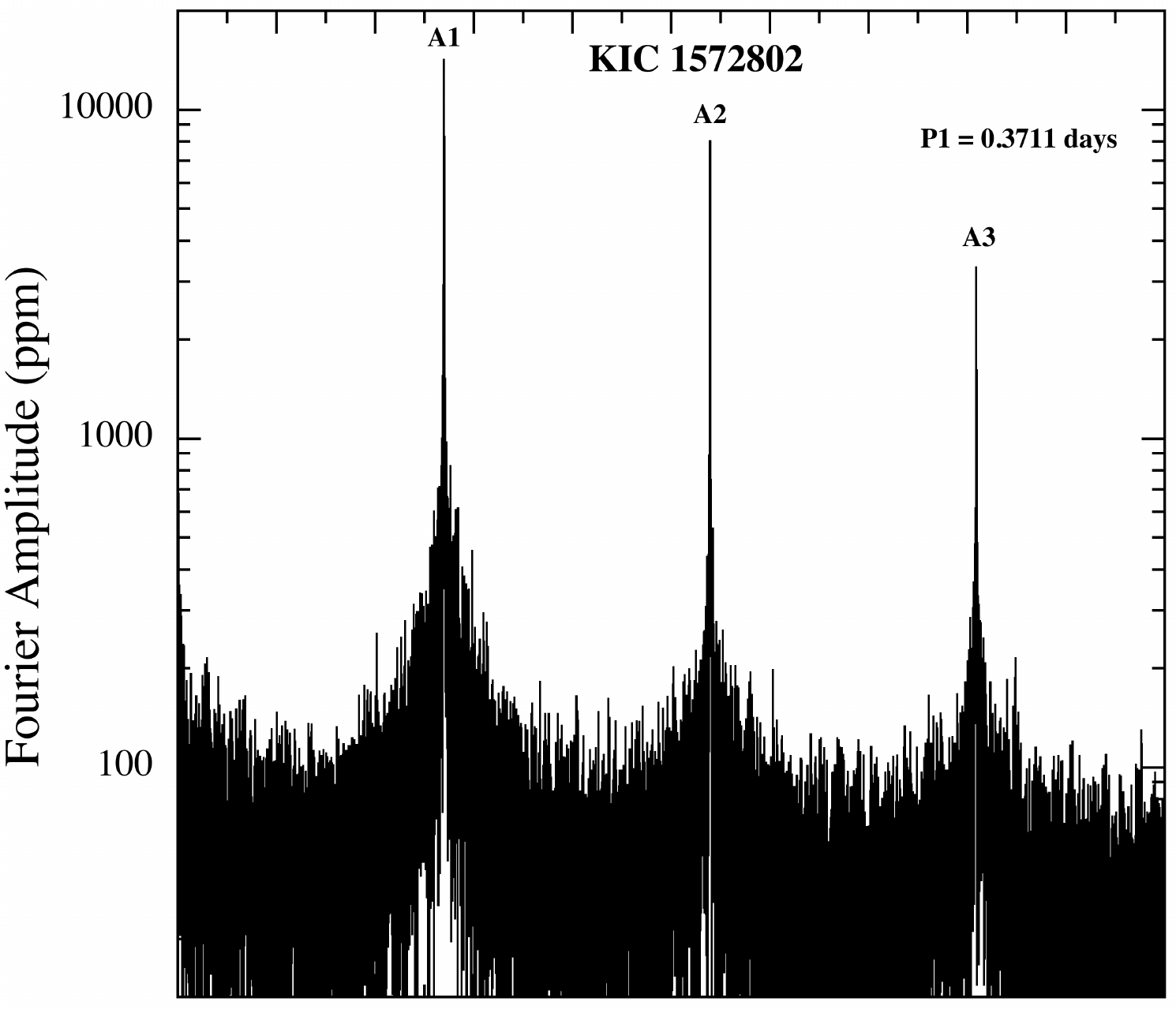} \hglue0.05cm
\includegraphics[width=0.424 \textwidth]{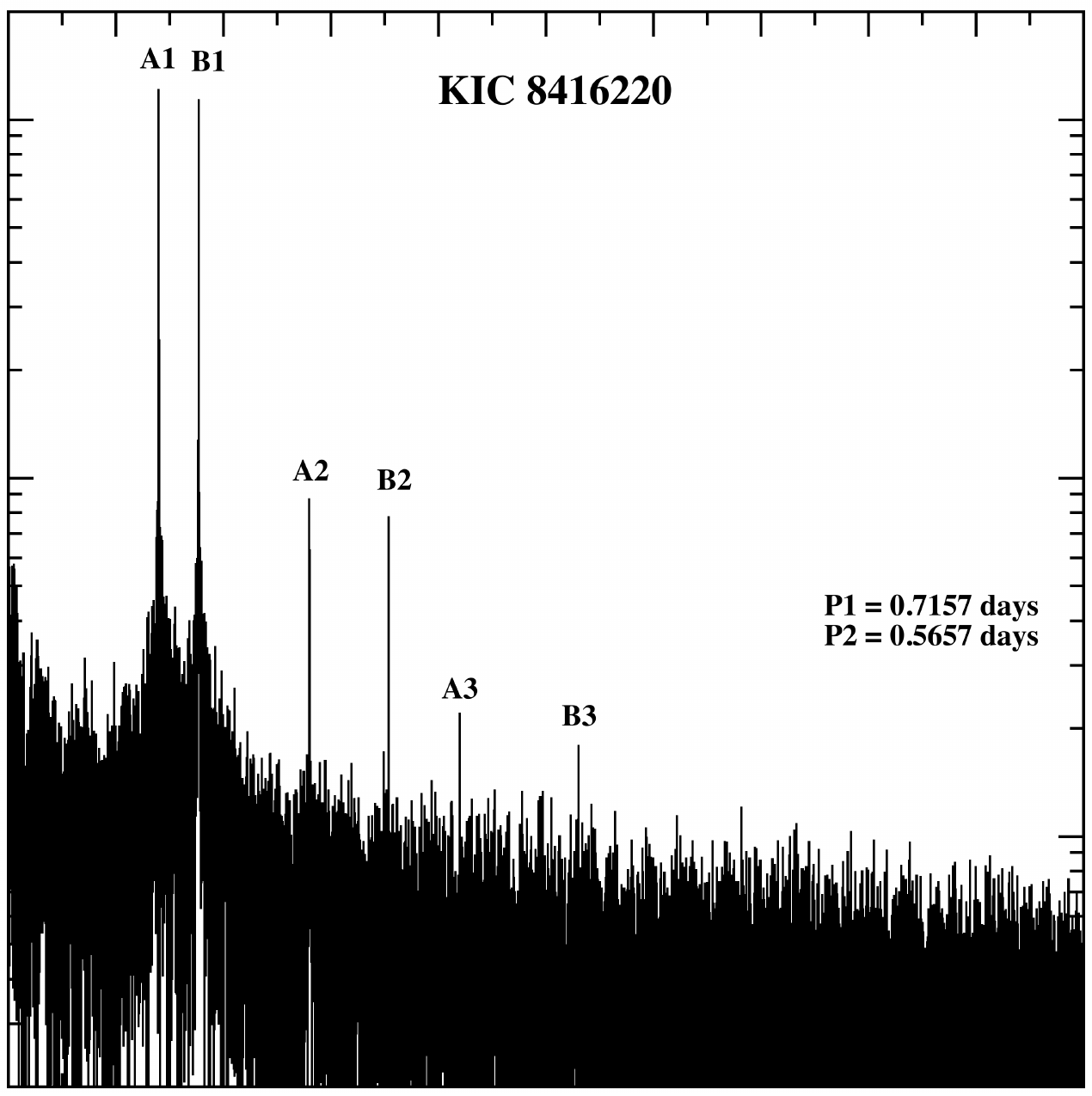} \vglue0.05cm
\includegraphics[width=0.496 \textwidth]{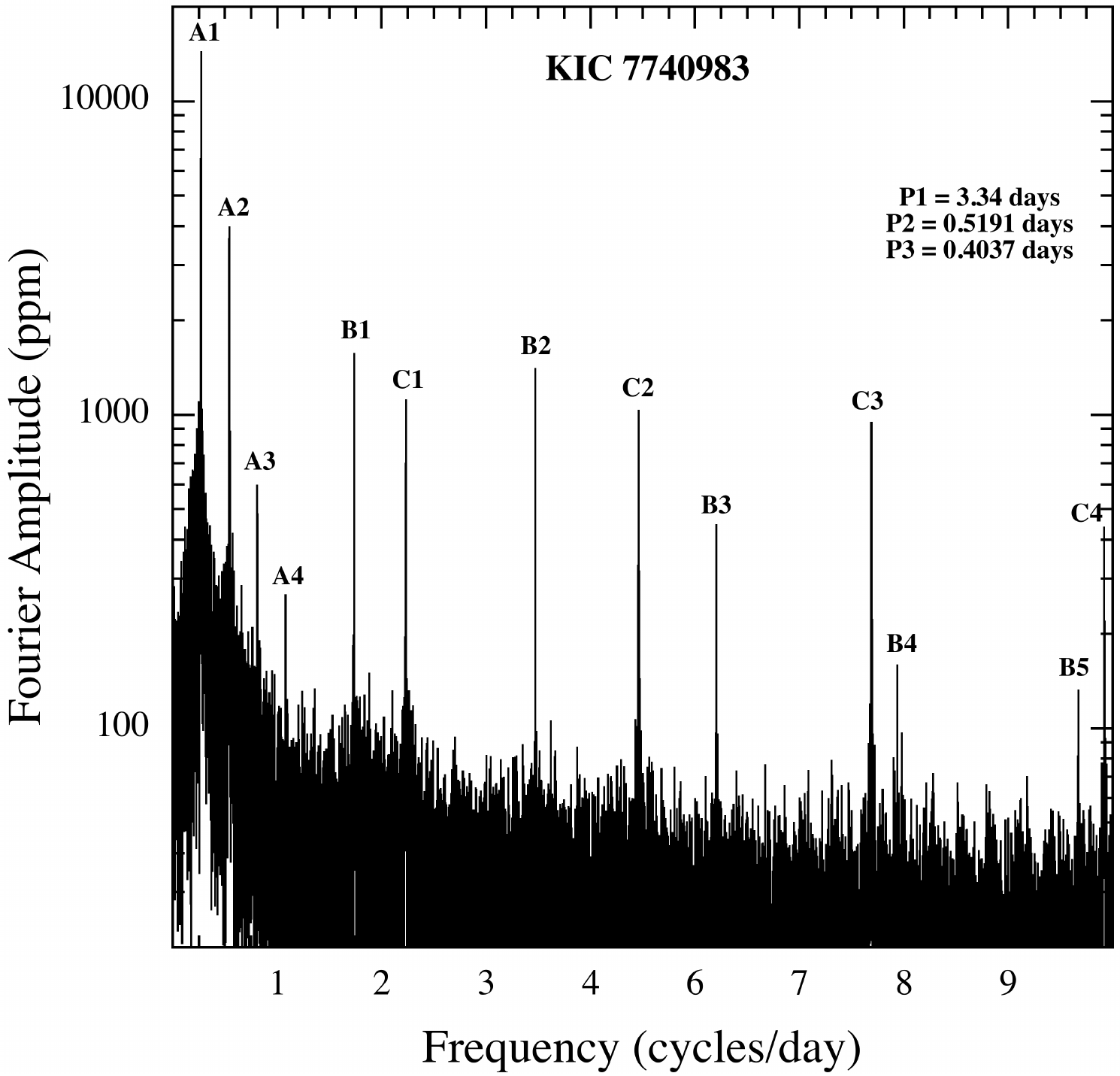} \hglue0.02cm
\includegraphics[width=0.425 \textwidth]{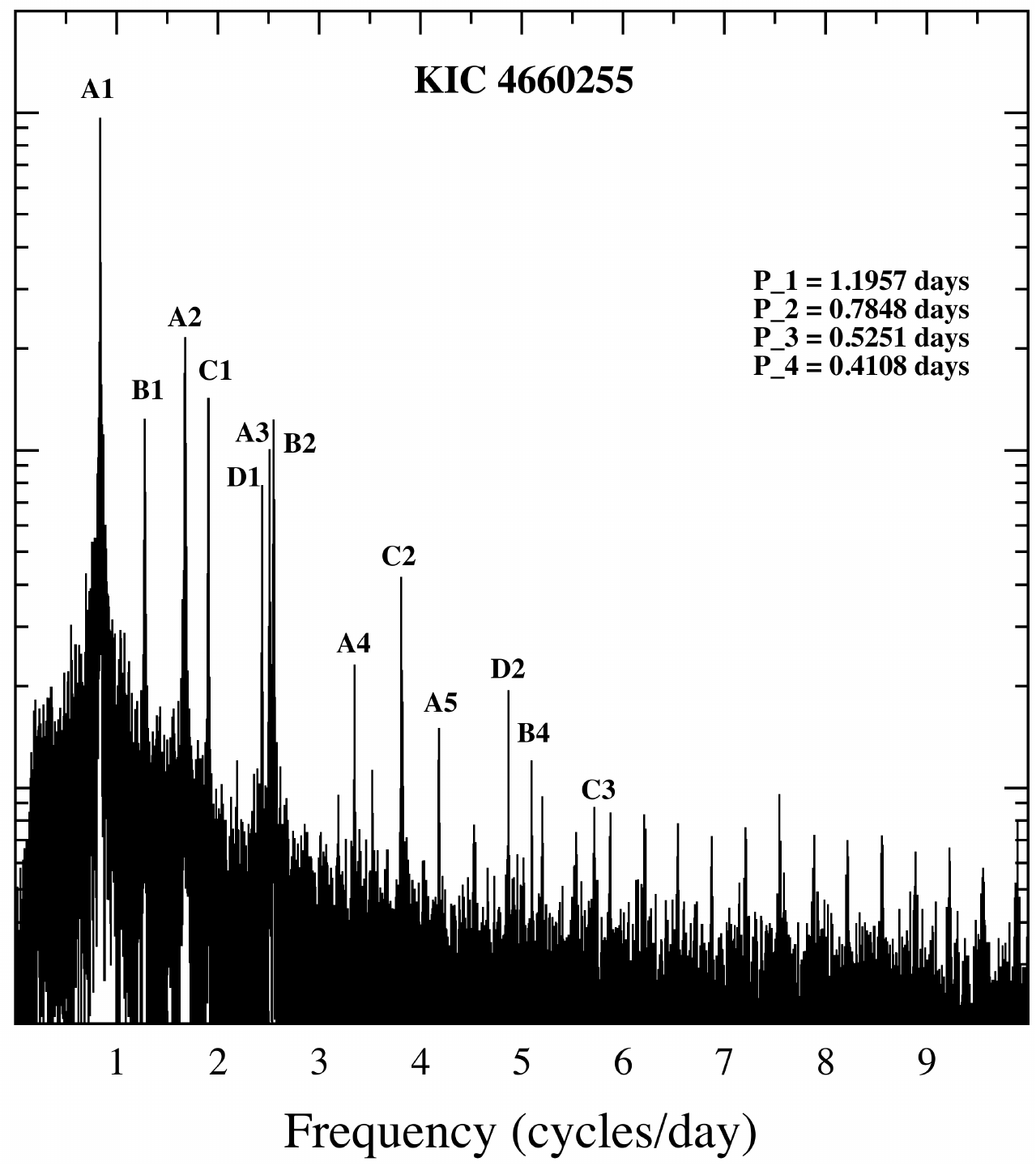} 
\caption{Four illustrative Fourier transforms of {\em Kepler} M-star targets.  KIC 1572802 has a single short rotation period; KIC 8416220 has two periods; KIC 7740983 has three; and KIC 4660255 has four rotation periods of $<$ 1 day.  The letters near the peaks designate one of the different periods, while the number which follows indicates the harmonic number -- starting with 1 for the base frequency.  The sequence of peaks in KIC 4660255 above frequencies of 6 cycles day$^{-1}$ and spaced by 1/3 cycle day$^{-1}$ is an artifact of the {\em Kepler} momentum wheel.}
\label{fig:FTs}
\end{center}
\end{figure*}

In all, we find 297 of the 3897 targets exhibit the requisite significant FT 
signal comprising a base frequency plus its harmonic, with the base frequency 
exceeding 0.5 cycles/day (i.e., $P_{\rm rot} < 2$ days).  We believe that the 
majority of these periodicities are likely to be due to stellar rotation manifested 
via starspots, but a significant number may be due to planet transits and binary 
eclipses. The individual FTs for these systems were further examined to
eliminate those which were clearly not due to rotating starspots.  In
all cases we folded the data modulo the detected fundamental period, and were
readily able to rule out cases due to transiting planets since their well-known
sharp, relatively rectangular dipping profiles are characteristic.  Of course, we also
checked the KOI list for matches.  Any of the objects
that appear in the {\em Kepler} eclipsing binary (``EB'') star catalog
(e.g., Matijevi\v c et al.~2012) were likewise eliminated.

Spots on a rotating star can, depending on their locations and other
properties, produce a profile that may be mistaken for the profile of an
EB.  The target KIC 1572802 is an example; see the top panel of
Fig.~\ref{fig:LCz}. 
It exhibits variations at a period of 8.9 hours
that resemble the light curve of a contact binary.  However, an
inspection of the top panel in Fig.~\ref{fig:LC}, which shows the 
corresponding full {\em Kepler} light curve, indicates that the object
is not likely to be a contact binary.  Aside from the profusion of stellar 
flares, which are not atypical of M stars, the overall envelope of the
8.9-hour periodicity is seen to be highly and erratically variable.
These amplitude changes are inconsistent with the behavior of a
typical EB.  Moreover, the shape of the 8.9-hr modulation changes
dramatically over the four years of observations.  The Fourier
transform for this object is shown in the top left panel of
Fig.~\ref{fig:FTs}.

The bottom panels of Figs.~\ref{fig:LC} and \ref{fig:LCz} provide
another example, i.e., the M star KIC 8416220 that exhibits a non-EB like
light curve, yet where the overall amplitude remains
nearly constant over the long term.  This light curve results from a beat
between two rotation periodicities whose periods differ by only $\sim$20\% (see the FT
in the upper right panel of Fig.~\ref{fig:FTs}).

\begin{figure*}
\begin{center}
\includegraphics[width=0.45 \textwidth]{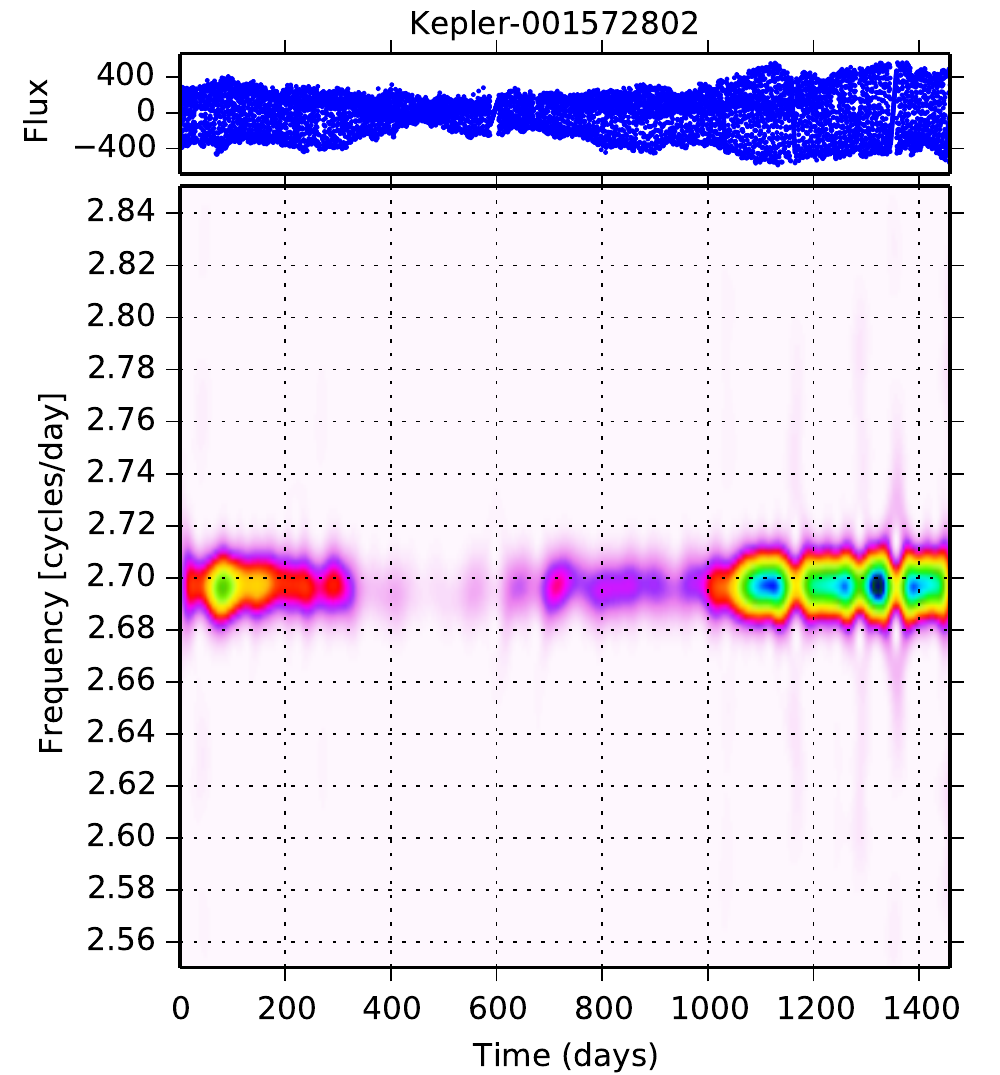} \hglue0.01cm
\includegraphics[width=0.45 \textwidth]{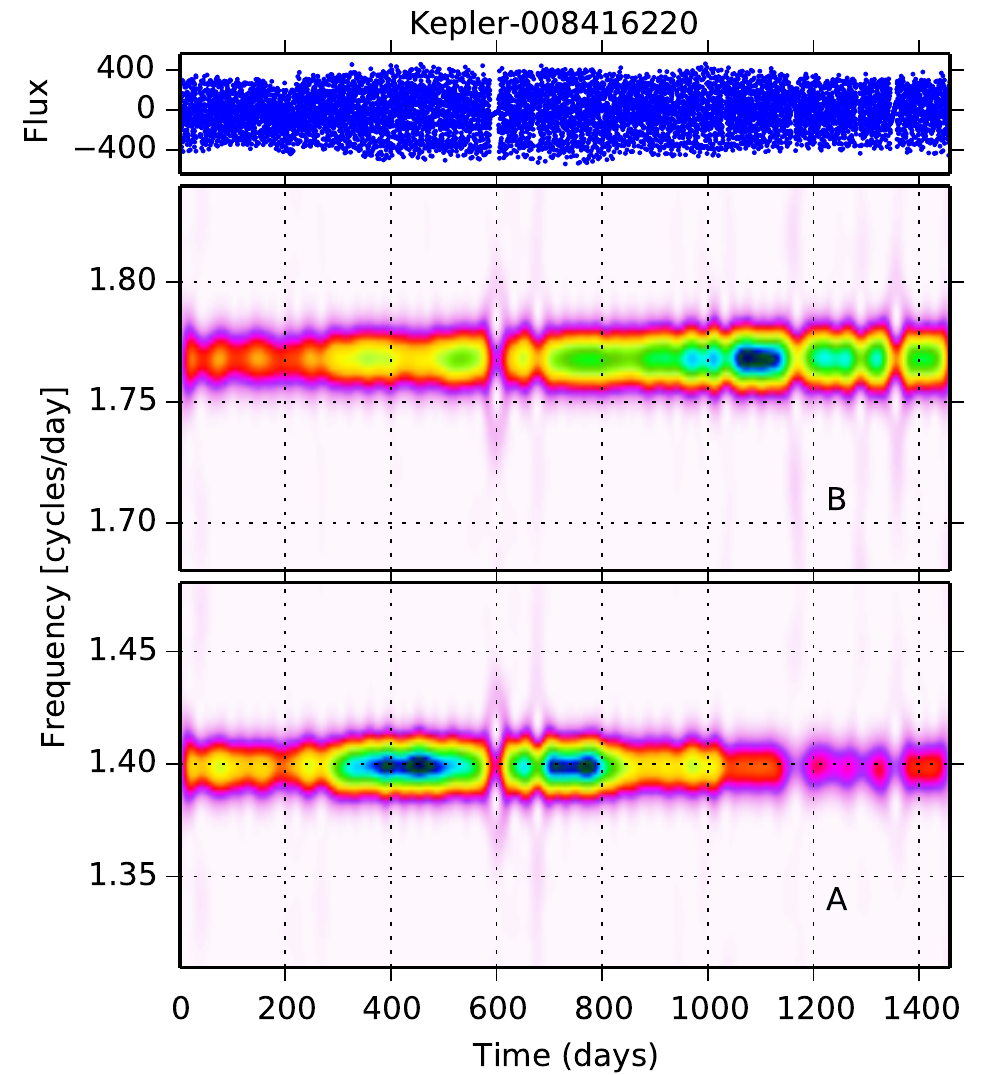} \vglue0.02cm
\includegraphics[width=0.45 \textwidth]{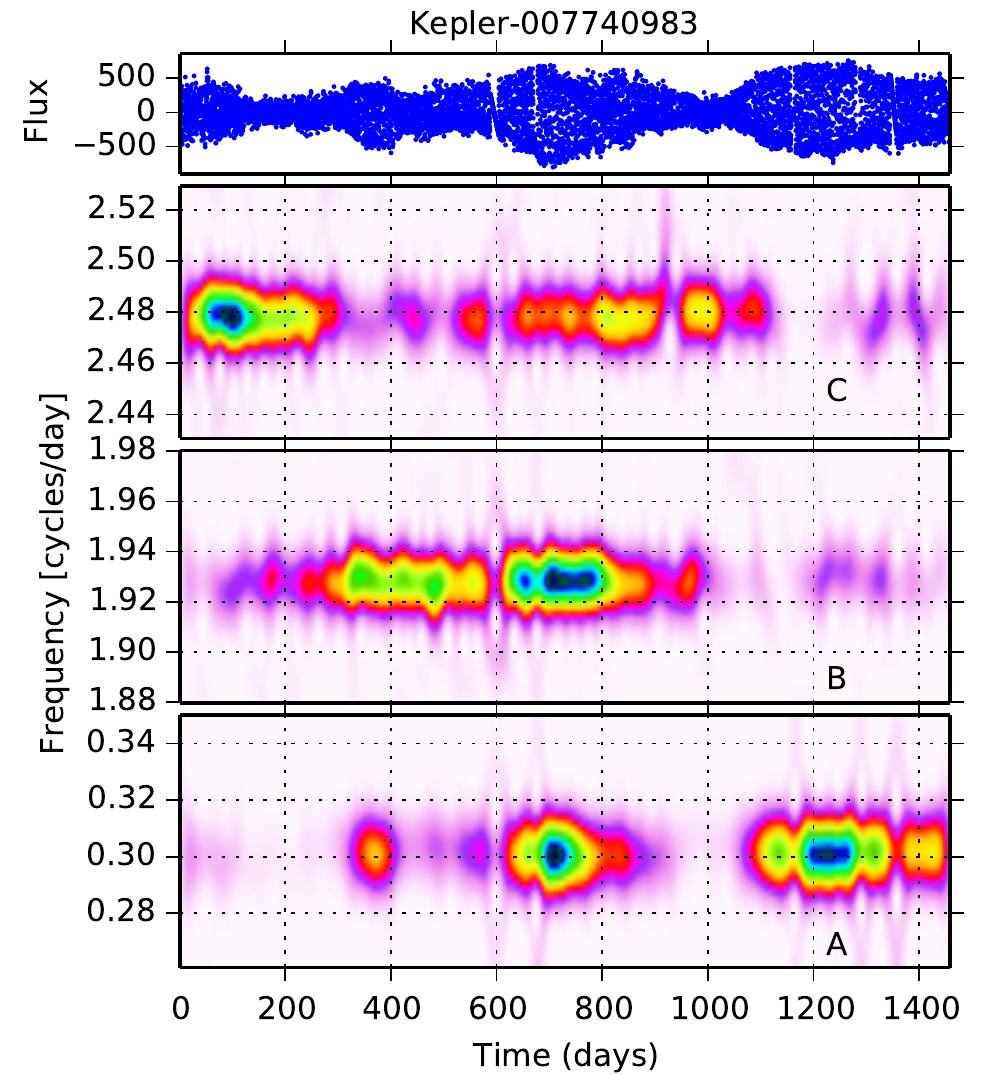} \hglue0.01cm
\includegraphics[width=0.45 \textwidth]{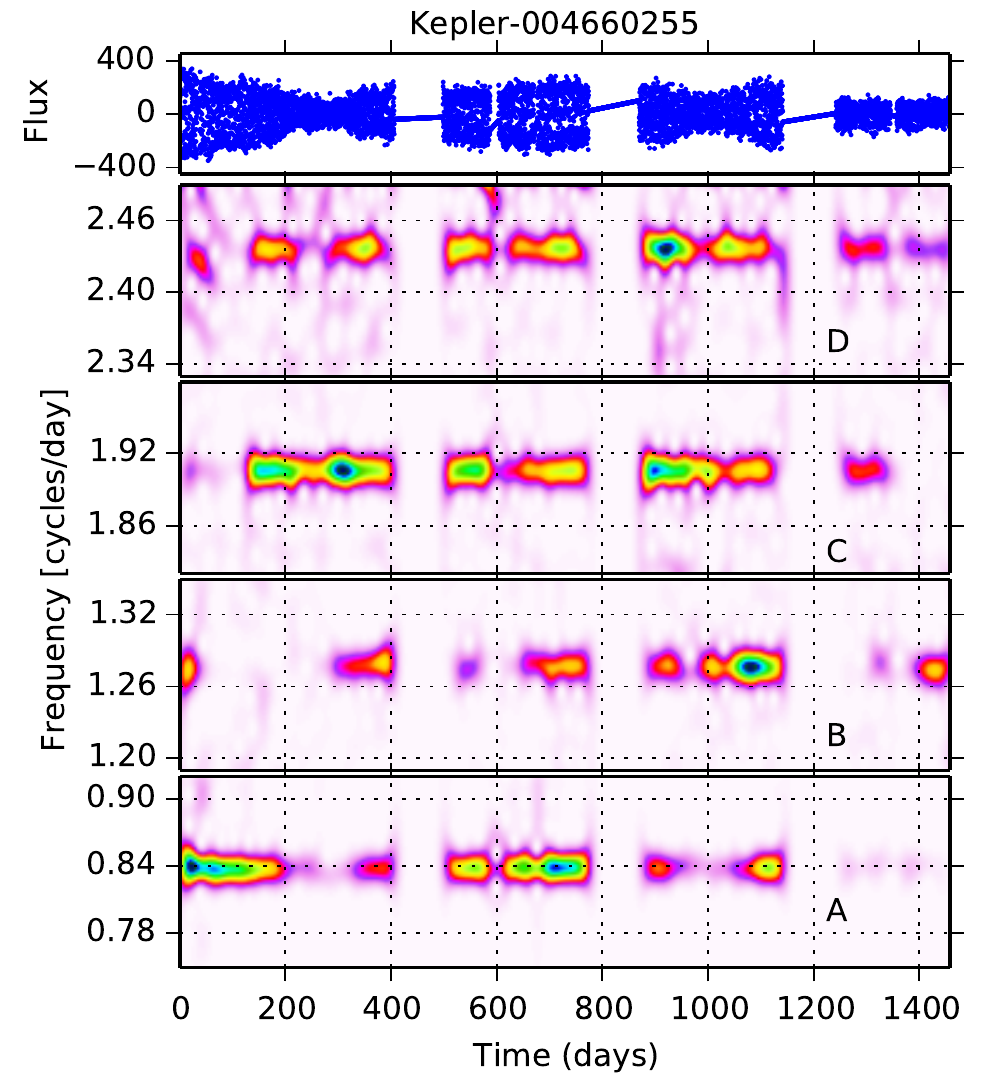} 
\caption{Illustrative ``sonograms'', or temporally resolved Fourier transforms, of the same {\em Kepler} M-star targets whose full FTs are shown in Fig.~\ref{fig:FTs}.  In each quadrant, the upper sub-panel shows the flux time series for the object, while the lower sub-panel is the sonogram.   
Each periodicity has its own intensity normalization.  The color coding is related to the FT amplitude with purple/red through green/blue the lower-to-higher amplitudes, respectively.  Note how most of the modulation amplitudes vary strongly with time and independently of one another.  These sonograms help demonstrate that the modulations are due to rotating starspots and not binary star modulations or stellar pulsations. }
\label{fig:sonogram}
\end{center}
\end{figure*}

After carefully examining the 297 M star targets that exhibit significant periodicities 
with periods less than two days, we believe we have identified the stellar rotators,
and are able to exclude the eclipsing binaries and transiting planets.  In summary,
we find that 
approximately 110 of these are previously identified binaries or transiting planets, 
or are likely to be binaries or transiting planets based on the characteristics of their 
FTs or folded light curves.  Some eight of the 297 targets were identified as artifacts 
caused by leakage of the signal from the very bright star RR Lyr that pulsates with a 
period of 0.5669 days.  After elimination of these targets, there remain 178 
targets that exhibit one or more periodicities that are likely to be due to starspots
(see Table \ref{tab:1Period}) on rotating stars.  These targets comprise a total
of 211 different periodicities with $P_{\rm rot} \lesssim 2$ days.
Some 110 of the 178 targets exhibit 127 periodicities 
with $P_{\rm rot} \lesssim 1$ day.  The distribution of all 211 periods is shown in the bottom
panel of Fig.~\ref{fig:periods}.

\section{Possible Stellar Pulsations Among the Periodicities}
\label{sec:pulsations}

Stars with convective envelopes oscillate in multiple modes that are
typically manifest as a forest of peaks in a Fourier spectrum, with
amplitudes forming a roughly Gaussian-shaped envelope (see, e.g.,
Bedding 2011 for a recent review). The frequency of the mode with the
maximum amplitude, $\nu_{\rm max}$, has been proposed to scale with
the acoustic cut-off frequency (Brown et al.~1991) and can be
related to basic stellar parameters by the scaling relation (Kjeldsen
\& Bedding 1995):
\begin{equation}
  \frac{\nu_{max}}{\nu_{max,\odot}}  =    \frac{M/M_{\odot}}{(R/R_{\odot})^2\sqrt{(T_{\rm eff}/5777K)}}
\end{equation}
Using the masses, effective temperatures, and radii of {\em Kepler} M-star targets as 
determined by Dressing \& Charbonneau (2013), as well as 
$\nu_{\rm max,{\odot}} \simeq$ 3140 $\mu$Hz (Barban et al.~2013), this relation yields $\nu_{\rm max}$ values in the 
range 6.4 - 31.6 mHz, corresponding to periods between 0.5 and 2.6 minutes. 
This is at least a factor of 80 shorter than the shortest period in our 
stellar sample. Additionally, scaling relations for amplitudes of convectively
excited oscillations predict amplitudes lower than a few parts per million for M dwarfs 
(Corsaro et al.~2013), which is incompatible with our observations. This, in 
combination with the lack of the typical Gaussian-shaped forests of peaks in 
our Fourier transforms, argues strongly against solar-like oscillations as the cause 
of the observed variability.

On the other hand, pulsations driven by the $\epsilon$ mechanism and
convective-flux blocking similar to those in $\gamma$ Doradus
variables have been theoretically predicted to occur in M dwarfs
(Rodr\'iguez-L\'opez et al.~2012, 2013). Usually, the associated
pulsation periods would be of the order of half an hour, but for some
low-mass pre-main sequence models timescales between 7 - 11 hours were
found.  However, the authors argued that these oscillations would only
grow during the short phase of deuterium burning, and, furthermore,
would only grow slowly during that phase.  It thus seems unlikely that
pulsations driven by this mechanism could be responsible for the 
observed periodicities. Further evidence against 
contamination of our sample by stellar pulsations is provided in Sections 
\ref{sec:TFA} and \ref{sec:phases}.

\section{Time-Frequency (``Sonogram'') Analysis}
\label{sec:TFA}

As a further check as to whether the periodicities in the {\em Kepler}
M-star data are the result of the rotation of spotted stars, we have
also carried out sonogram, i.e., time-frequency, analyses.  In the
present case, short segments of the {\em Kepler} time series are
Fourier transformed and the amplitudes in each FT are
displayed as a function of frequency in a vertical linear strip of a
two-dimensional image.  The orthogonal coordinate of the image encodes
the start times of the data segments. The duration of the individual
segments of the time series, and their overlap are parameters that can
be adjusted. The image shows how the amplitude of a particular
signal evolves with time.

To compute the sonograms, we used the program package {\tt TiFrAn}
(Time Frequency Analysis; Koll{\'a}th \& Ol{\'a}h 2009) which allows a
time-frequency analysis to be carried out via several different
methods.  From these possible options, we have chosen the Short-Term
Fourier Transform (STFT) to study the rotational periodicities. This
procedure yields good resolution in both time and frequency. 
In particular, each data segment is defined by a Gaussian window function 
with full width at half maximum (FWHM) of about 30 days, and the centers 
of the windows of sequential segments are separated by $\sim$2 days.

\begin{figure*}
\begin{center}
\includegraphics[width=0.45 \textwidth]{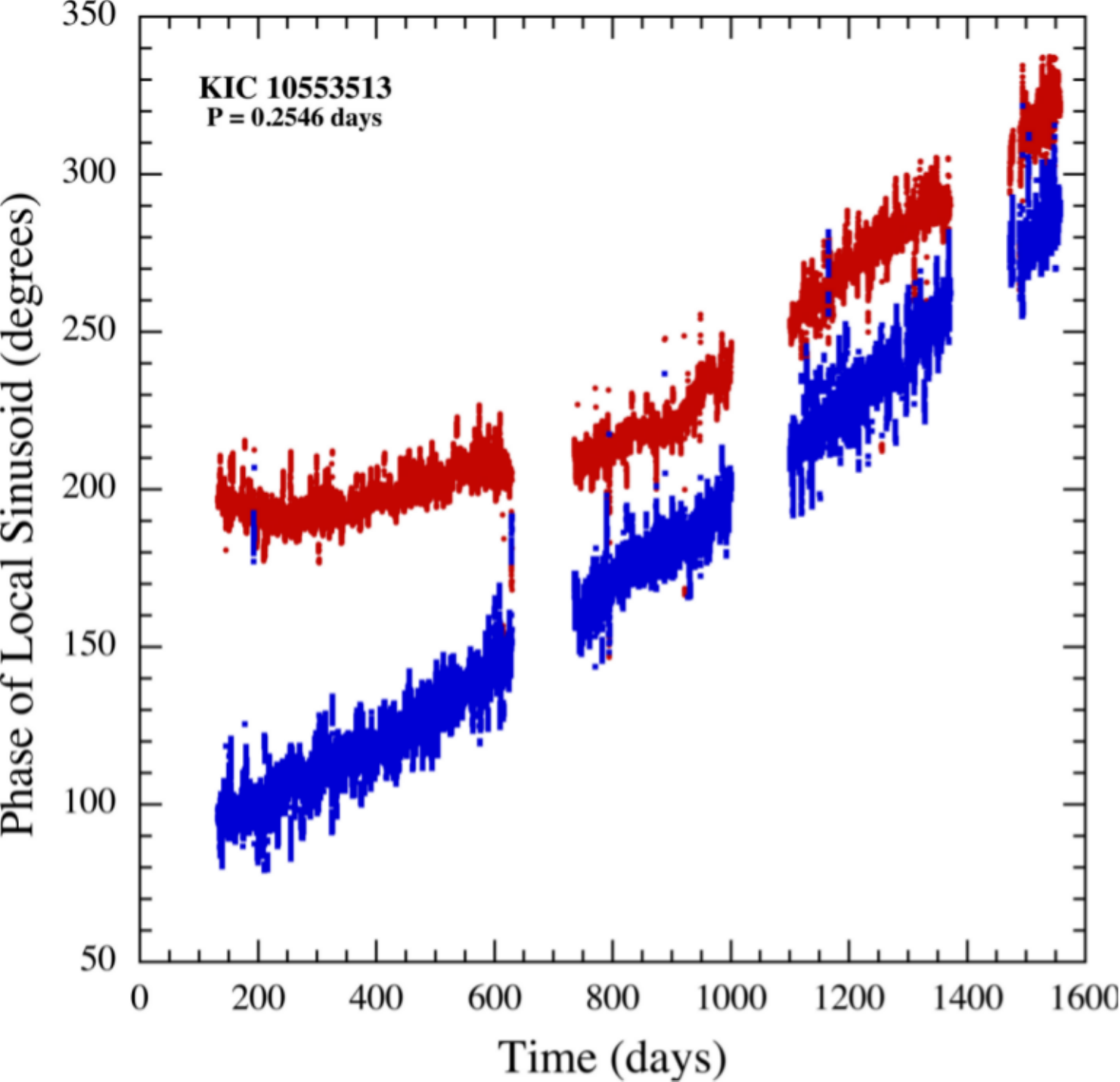} \hglue0.05cm
\includegraphics[width=0.45 \textwidth]{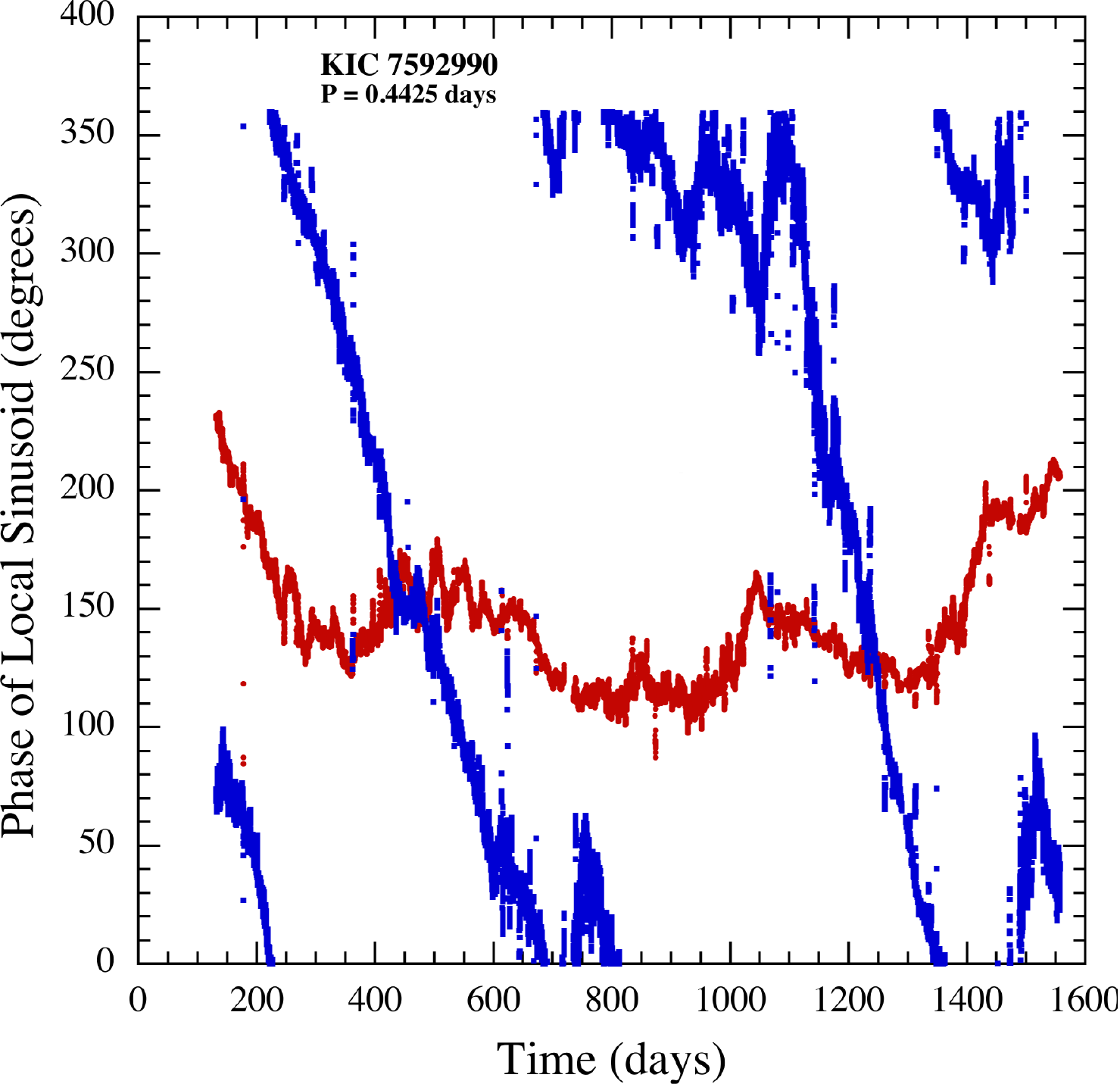} \vglue0.05cm
\includegraphics[width=0.45 \textwidth]{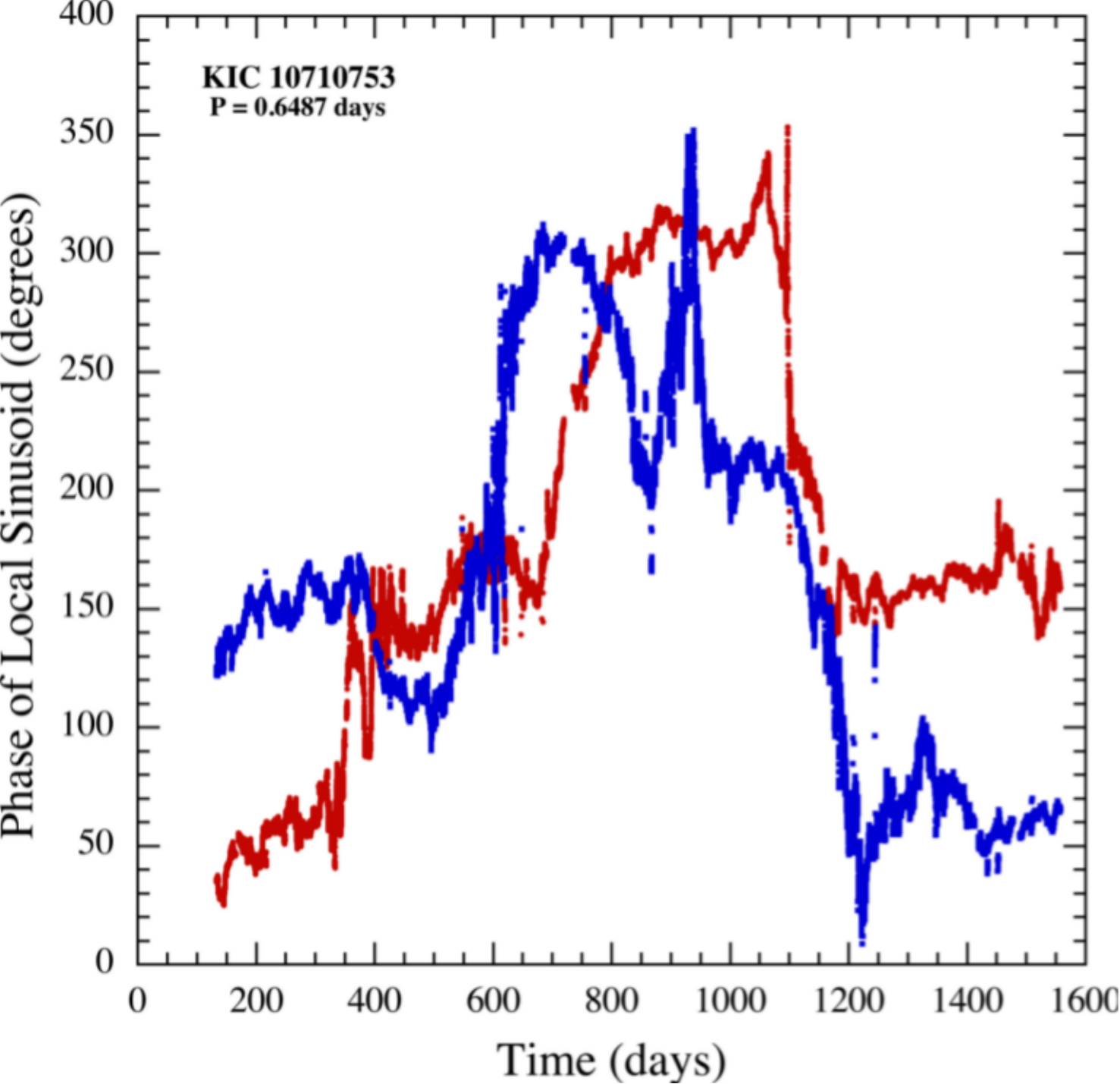} \hglue0.05cm
\includegraphics[width=0.45 \textwidth]{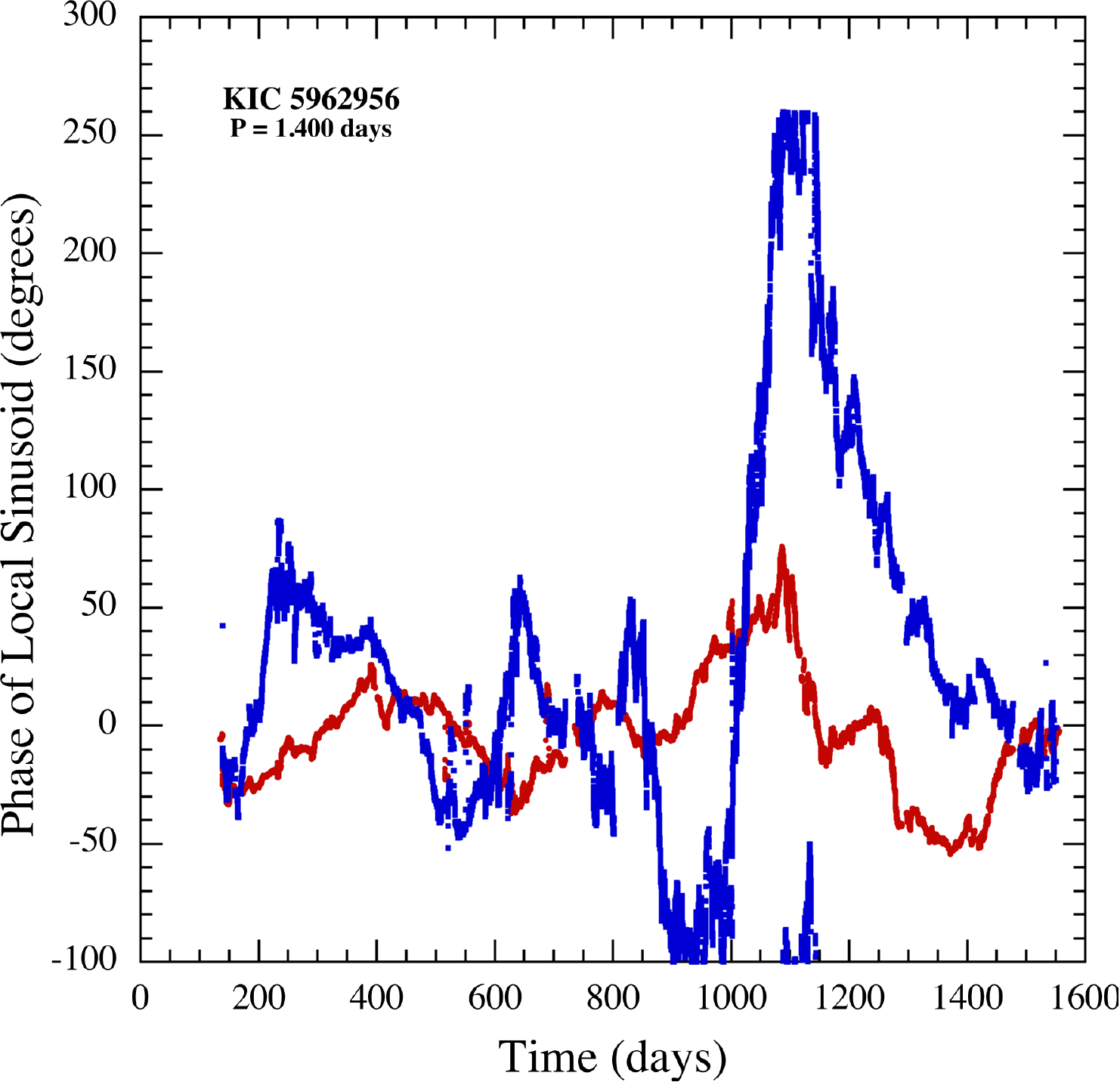} 
\caption{Illustrative phase tracking diagrams for four systems that appear to exhibit periodic modulations due to starspot rotations. The inferred rotation periods range from 0.25 days to 1.4 days.  The red and blue curves are the fitted values with time of $\phi_1$ and $\phi_2$ defined in Eqn.~(\ref{eqn:phz}), which represent the locally defined phase of the base frequency and its first harmonic, respectively.  The two phases are typically seen to vary erratically with time, as might be expected for transitory, migrating, and differentially rotating starspots.}
\label{fig:phases}
\end{center}
\end{figure*}

In Fig.~\ref{fig:sonogram} we show sonograms for the same four {\em
Kepler} targets whose full FTs are displayed in Fig.~\ref{fig:FTs}.
The FT amplitudes of most of the individual frequency components appear to 
change dramatically and erratically over time, and largely independently of one 
another.  This gives added weight to the hypothesis that these are indeed features 
due to spots on rotating stars, as opposed to stellar pulsations or other more
stable periodicities, e.g., binary modulations.

\section{Tracking the Modulation Phases}
\label{sec:phases}

It would not be surprising to find that starspots produce modulations 
with phases that are quite variable, as are the amplitudes per the 
above discussion. Therefore,
we devised a way to track the phases of the observed periodicities in
order to assess their characteristics and determine whether they 
suggest an origin in starspots.  Because the FTs of the inferred
rotationally induced spot modulations have harmonics that fall off
rapidly in amplitude with frequency, we model the modulations by just
the base frequency and its first harmonic.  We can then describe the
flux locally in the time series by the expression:
\begin{eqnarray}
\label{eqn:phz}
\mathcal{F}(t) = A + B \sin(\omega t + \phi_1) + C \sin(2\omega t + \phi_2)
\end{eqnarray}
where $\omega \equiv 2 \pi/P_{\rm rot}$.  For each point in the flux
time series, we fit a function of the form given by
Eqn.~(\ref{eqn:phz}) over a short interval of time equal to 2 to 4
base periods of the modulations, depending on the statistical
precision that is desired.  For periods between 1/4 and 2 days, this
corresponds to using between 25 and 400 flux points to determine five
unknown parameters, including the phases $\phi_1$ and $\phi_2$. The
phases of the base frequency and of its first harmonic are thereby
computed for every point in the time series.

Results of this `phase tracking' are shown in Fig.~\ref{fig:phases}
for four M-star targets for which we found, in each case, a single
rotation period in the range from 0.25 to 1.4 days.  The phases
of the base frequency are plotted in red points while those of the
first harmonic are shown in blue.  Portions of the phase curves with
linear trends indicate times of constant effective frequency; regions
with curvature indicate changes in the apparent frequency.  These
plots tend to exhibit two different signatures that are likely
characteristic of transitory, migrating, and differentially rotating
starspots: (1) erratic phase behavior, and (2) different phase
behavior for the base frequency vs.~the first harmonic.  
The stellar pulsators that we have examined with this same phase-tracking
code do not behave in this manner.

More quantitatively, the phase of the modulations can be formally defined as
\begin{eqnarray}
\phi(t) = \int (\nu(t) - \nu_0) \, dt~~{\rm or}~~~ \dot \phi(t) = \nu(t) - \nu_0
\end{eqnarray}
In terms of the period, this can be written as
\begin{eqnarray}
\label{eqn:dP}
P(t) \simeq P_0 - \dot \phi(t) \,P_0^2
\end{eqnarray}
where $\nu_0$ and $P_0$ are the reference frequency and period, respectively.  The 
characteristic timescale for period changes, $\tau$, can be expressed in terms of $\ddot \phi$ as:
\begin{eqnarray}
\tau \equiv \frac{P}{\dot P} \, \simeq \frac{1}{P_0 \ddot \phi} 
\end{eqnarray}
As numerical examples, we note that a change in phase of 1 cycle
(360$^\circ$) over the {\em Kepler mission} corresponds to $\dot \phi
\simeq$ (1500 d)$^{-1}$ which implies, according to
Eqn.~(\ref{eqn:dP}), a difference in period from the reference period
of $\sim$0.00017 days (for $P_0 = 1/2$ day).  In terms of the
implications of curvature in the phase curves, a parabolic arc of 1/6
of a cycle in amplitude occurring over an interval of $\pm 200$ days
corresponds to $\ddot \phi \simeq 8 \times 10^{-6}$ cycles day$^{-2}$ with a
corresponding value of $\tau \simeq 330$ years.

\begin{figure}
\begin{center}
\includegraphics[width=0.97 \columnwidth]{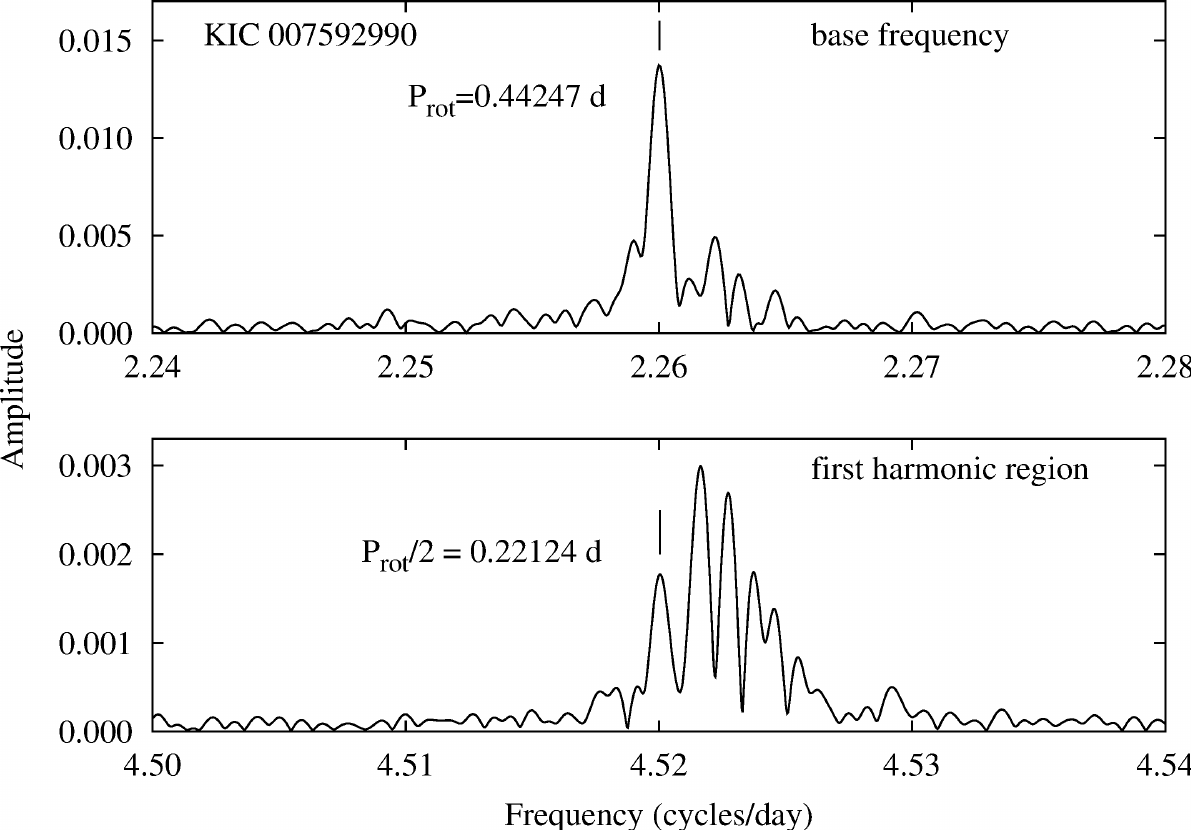} 
\caption{Zoom-in on the Fourier transform for KIC 7592990 around the base frequency and its first harmonic.  Note that both peaks are broadened well beyond the frequency resolution of the 1500-day {\em Kepler} data train.  The aligned vertical lines in the two panels represent the base frequency and twice the base frequency.  As noted in the text, it is evident that the centroid of the first harmonic has a frequency that is clearly higher than twice that of the base frequency.  This effect is also demonstrated by the phase plot in Fig.~\ref{fig:phases} and indicates starspots at different latitudes undergoing differential rotation.}
\label{fig:KIC759}
\end{center}
\end{figure}

Finally, we note an important property of these ``phase curves''.  By
construction from Eqn.~(\ref{eqn:phz}), we see that if the second
harmonic term is exactly twice the base frequency, then any slopes in
the phase curves should bear the relation $\dot \phi_2 = 2 \dot
\phi_1$.  However we can see examples in Fig.~\ref{fig:phases} where
this is clearly not the case.  For KIC 10553513 the two slopes are
nearly the same over the final 800 days of observation, while for KIC
7592990 the mean slope of $\phi_2$ is substantially larger than twice
the mean slope of $\phi_1$.  This is a direct indication that the
first harmonic does not occur at exactly twice the base frequency (see
also Fig.~\ref{fig:KIC759}). Vida \& Ol\'ah (2013) and Vida et al.~(2014) find 
that KIC 7592990 has an activity cycle on the timescale of about 520 days
inferred from the systematic change of its rotational period due to
differential rotation, with an estimated lower limit of
$\alpha=0.0012$ (see Eqn.~\ref{eqn:alpha}). In turn, this likely
demonstrates that the second harmonic may arise from spots that are
located at both different longitudes {\em and} different latitudes,
with the attendant differential rotation when latitude differences are present.  Clustering of
starspots at two active longitudes opposite to each other, on close
binaries of dwarf stars, has been theoretically investigated by
Holzwarth \& Sch\"ussler (2003), showing that with faster rotation the
occurrence of this clustering is higher (but the initial parameters of
the rising flux tubes also play a significant role).  

\section{Multiple Period Systems}
\label{sec:multiple}

Upon further careful examination of the FTs of the 178 systems with short
rotation periods among the {\em Kepler} M stars, we find $\sim$30 objects with 
two or more incommensurate periods, at least one of which has $P_{\rm rot} <$
1 day, and a second that has $P_{\rm rot} <$ 2 days.  KIC 8416220,
discussed above, provides an example of an M star that exhibits two
periodicities; see the upper right panel of Fig.~\ref{fig:FTs}.  Note
that the amplitudes of all three visible harmonics at each frequency
have very similar amplitudes.  The two base frequencies differ by only
$\sim$20\%. 

Further scrutiny also reveals several objects with three or more independent short 
periods, at least one of which has $P_{\rm rot} <$ 1 day.   One particular example 
is KIC 7740983 whose FT is shown in the lower left panel of Fig.~\ref{fig:FTs}.  The 
three independent frequencies, labeled ``$A$'', ``$B$'', and ``$C$'' exhibit 3, 7, and 
11 harmonics, respectively, out to 25 cycles/day.  

One of the systems whose FT we studied exhibited {\em four} independent frequencies: 
KIC 4660255 (see lower right panel of Fig.~\ref{fig:FTs}).  All four periods have $P_{\rm rot} <$1.2 days.

The results of the multiple M-star detections are summarized in Table
\ref{tab:3Periods} which lists 37 systems.  Of these, 27 have one
period with $P_{\rm rot} < 1$ day, another with $P_{\rm rot} < 2$
days, and each independent frequency has at least a base frequency
plus the next higher harmonic.  For 7 of the other systems listed in
Table \ref{tab:3Periods}, they satisfy all of these criteria, except
that one of the periodicities does not have a detectable harmonic.
Finally, a few
remaining systems have a second period that is slightly longer than 2
days.

\begin{figure*}
\begin{center}
\includegraphics[width=0.45 \textwidth]{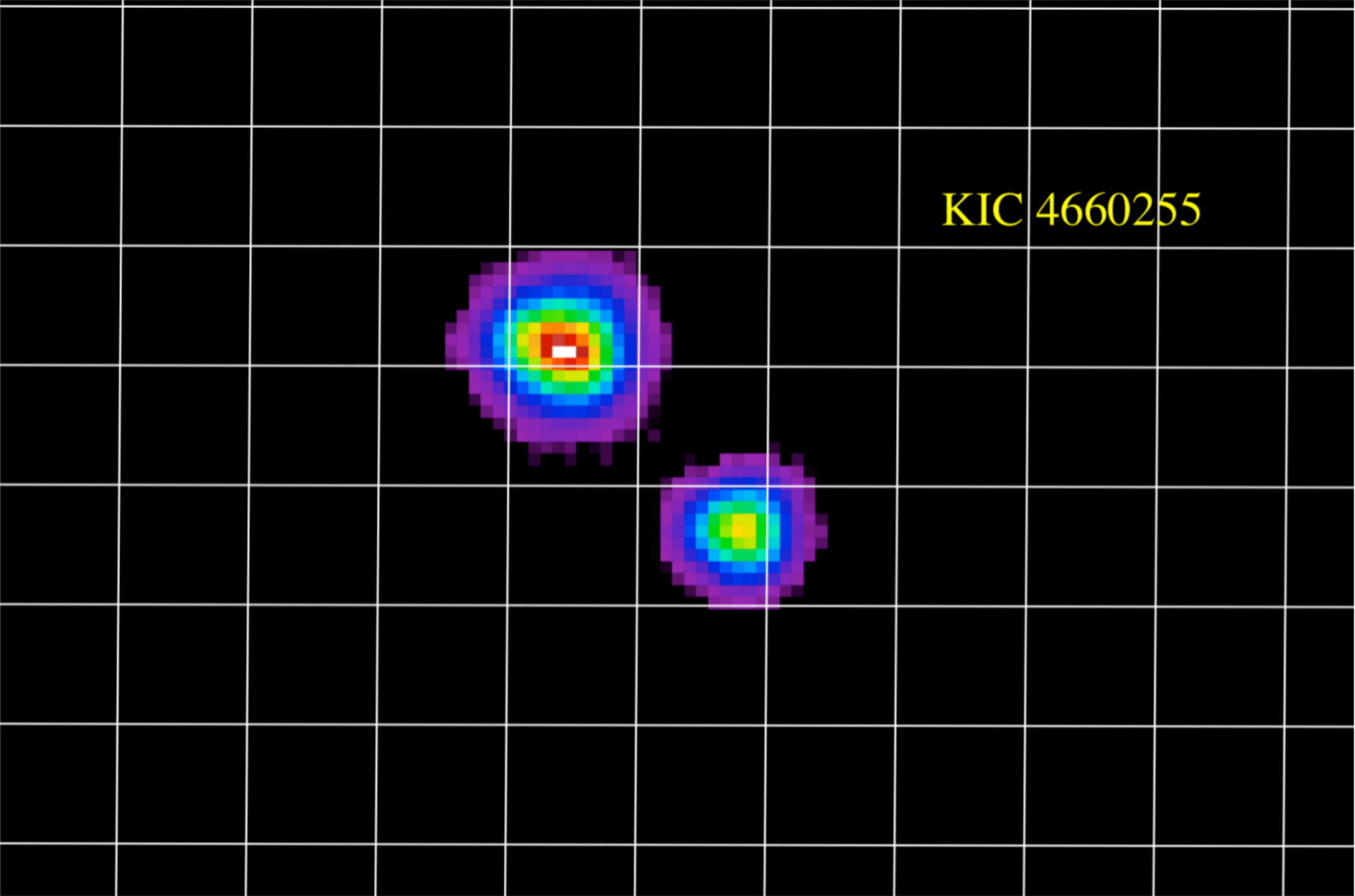} \hglue0.1cm
\includegraphics[width=0.45 \textwidth]{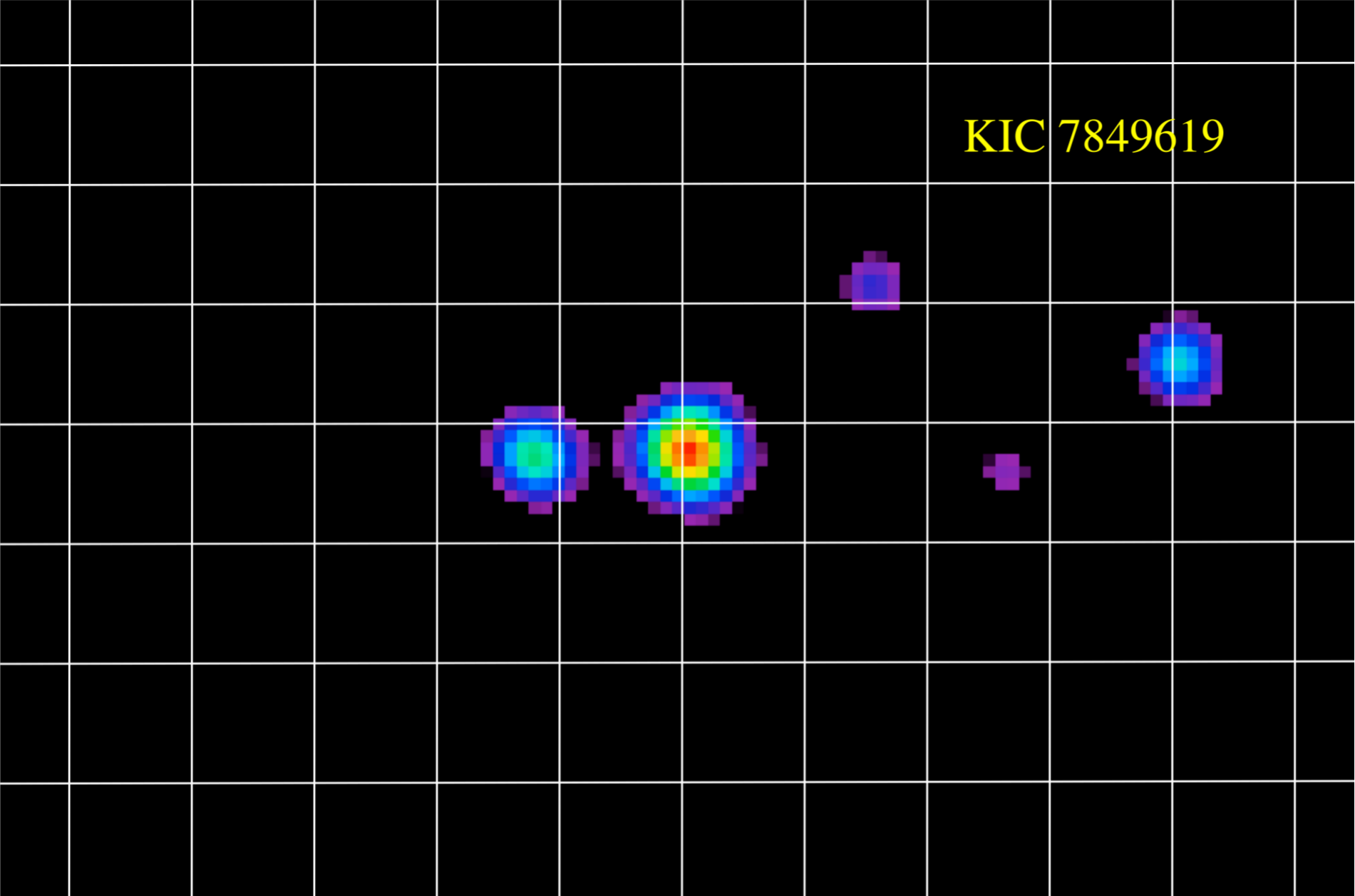} \vglue0.1cm
\includegraphics[width=0.45 \textwidth]{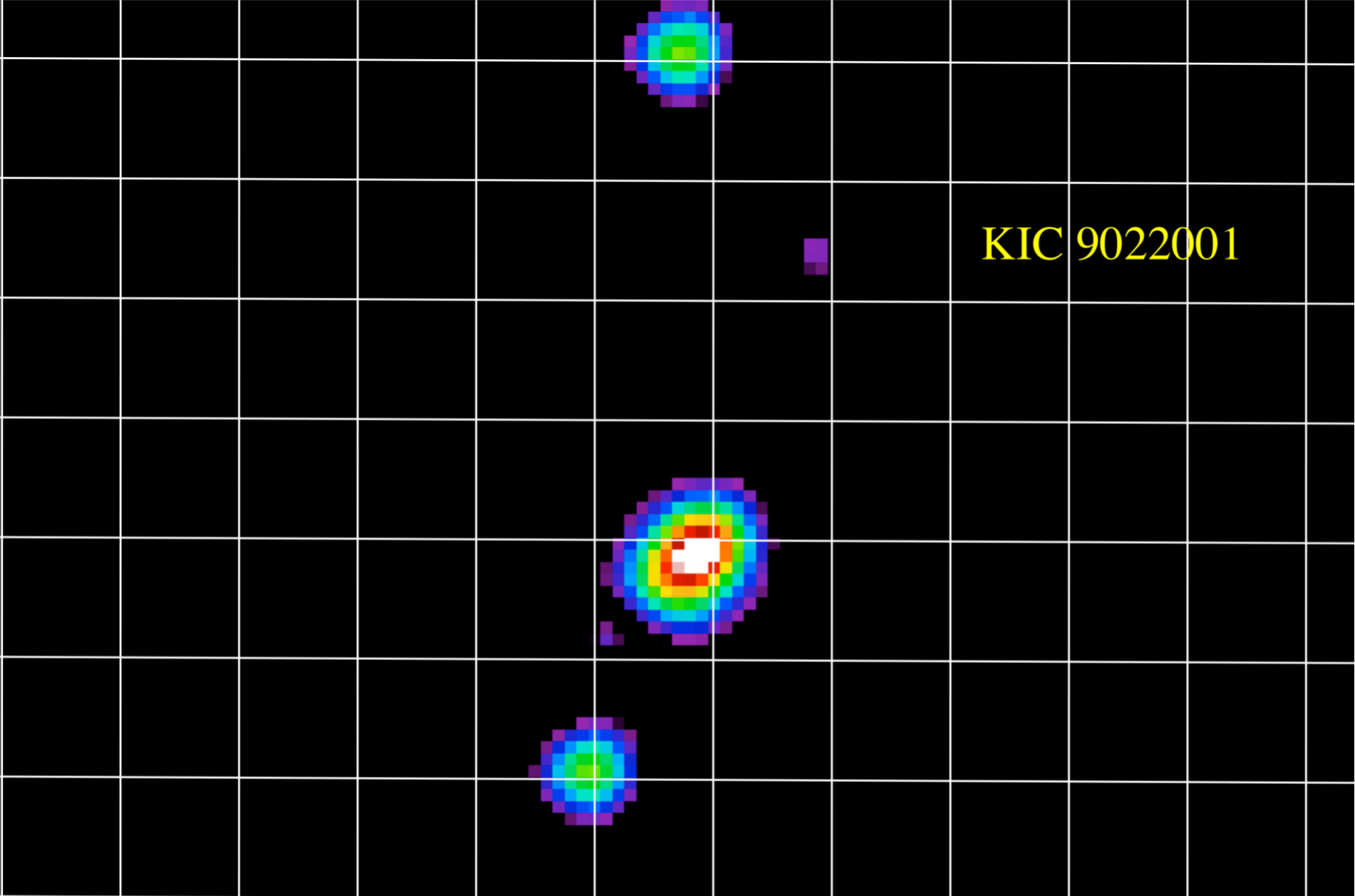} \hglue0.1cm
\includegraphics[width=0.45 \textwidth]{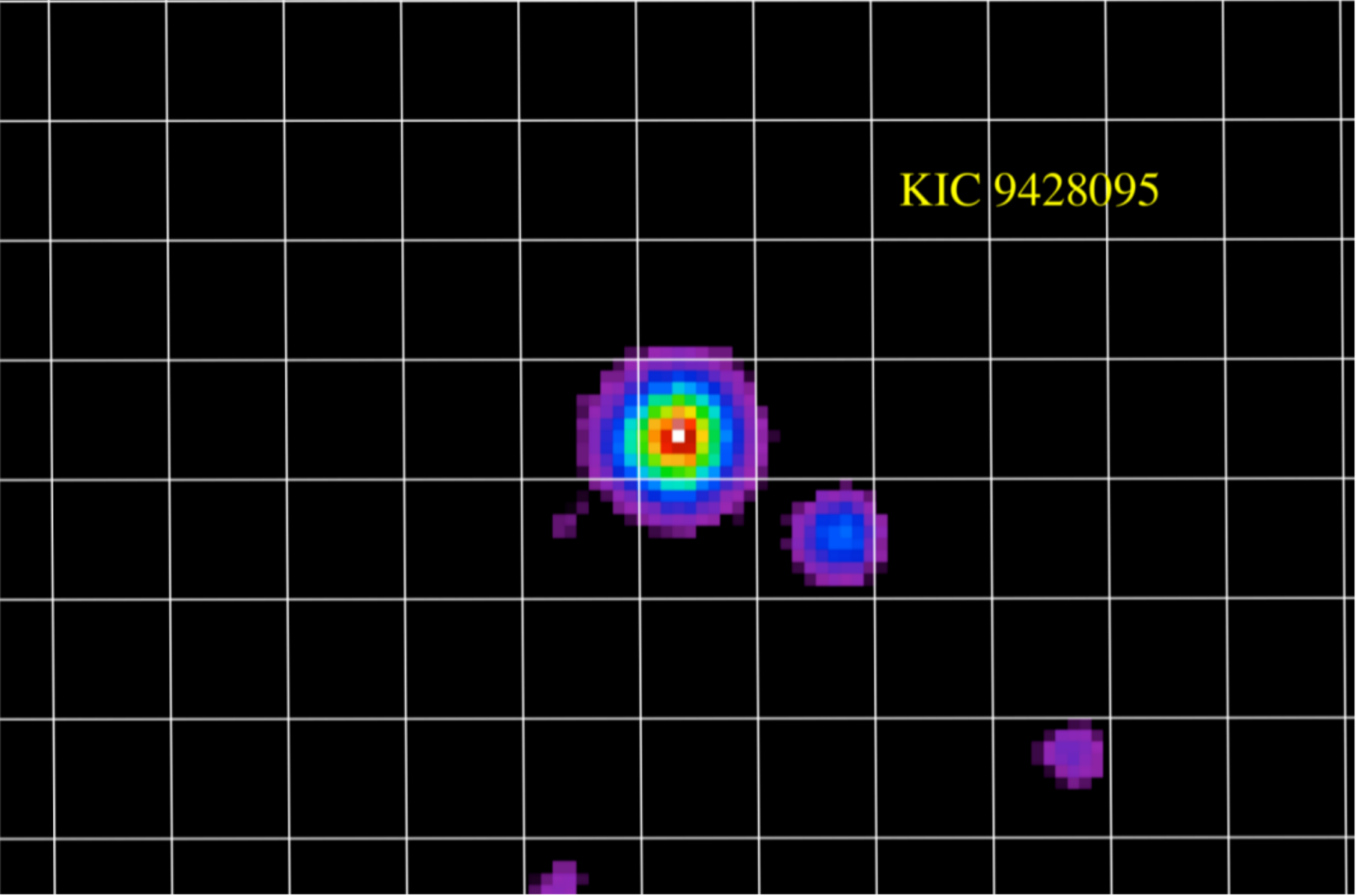} \vglue0.1cm
\includegraphics[width=0.45 \textwidth]{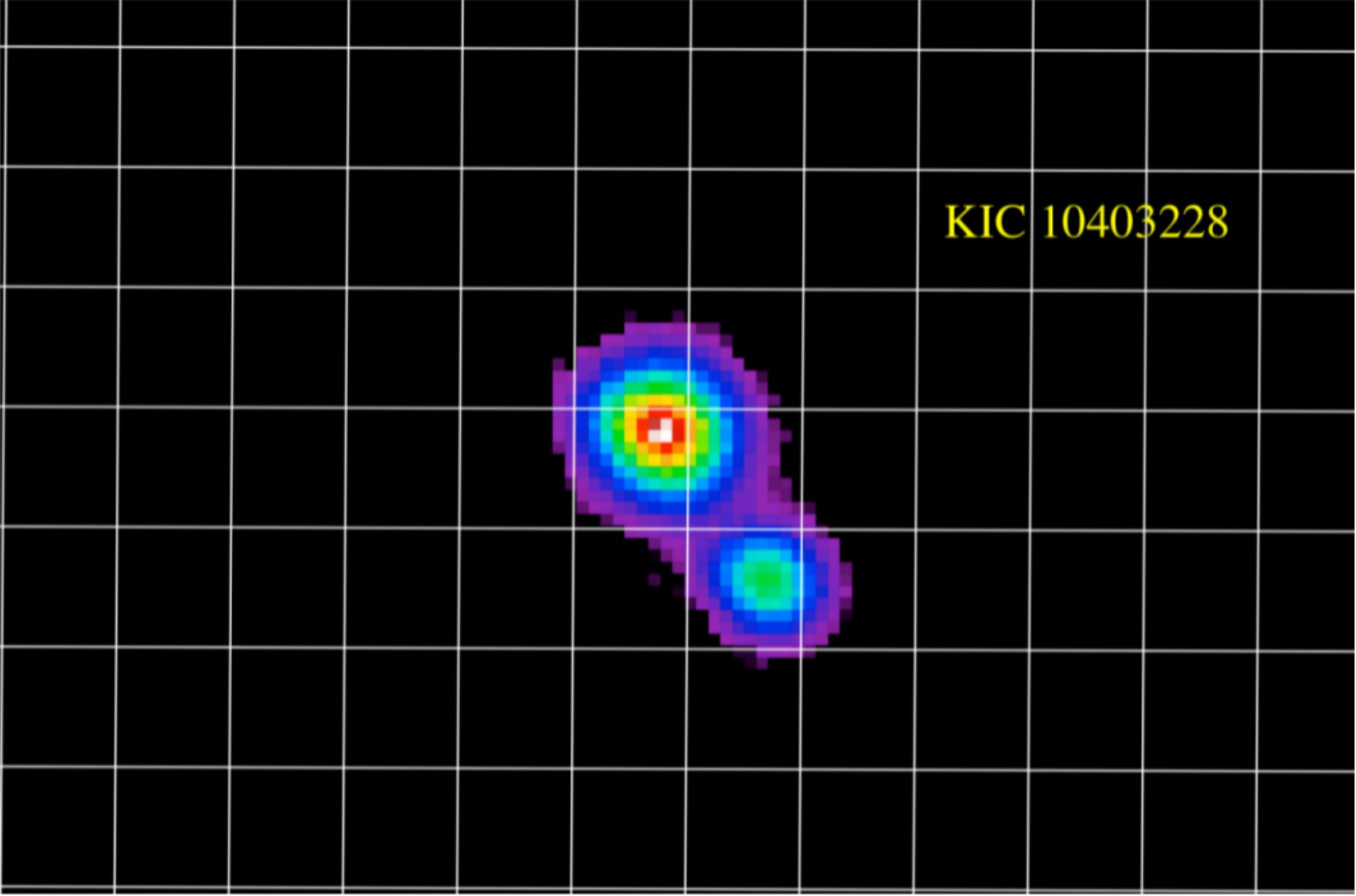} \hglue0.1cm
\includegraphics[width=0.45 \textwidth]{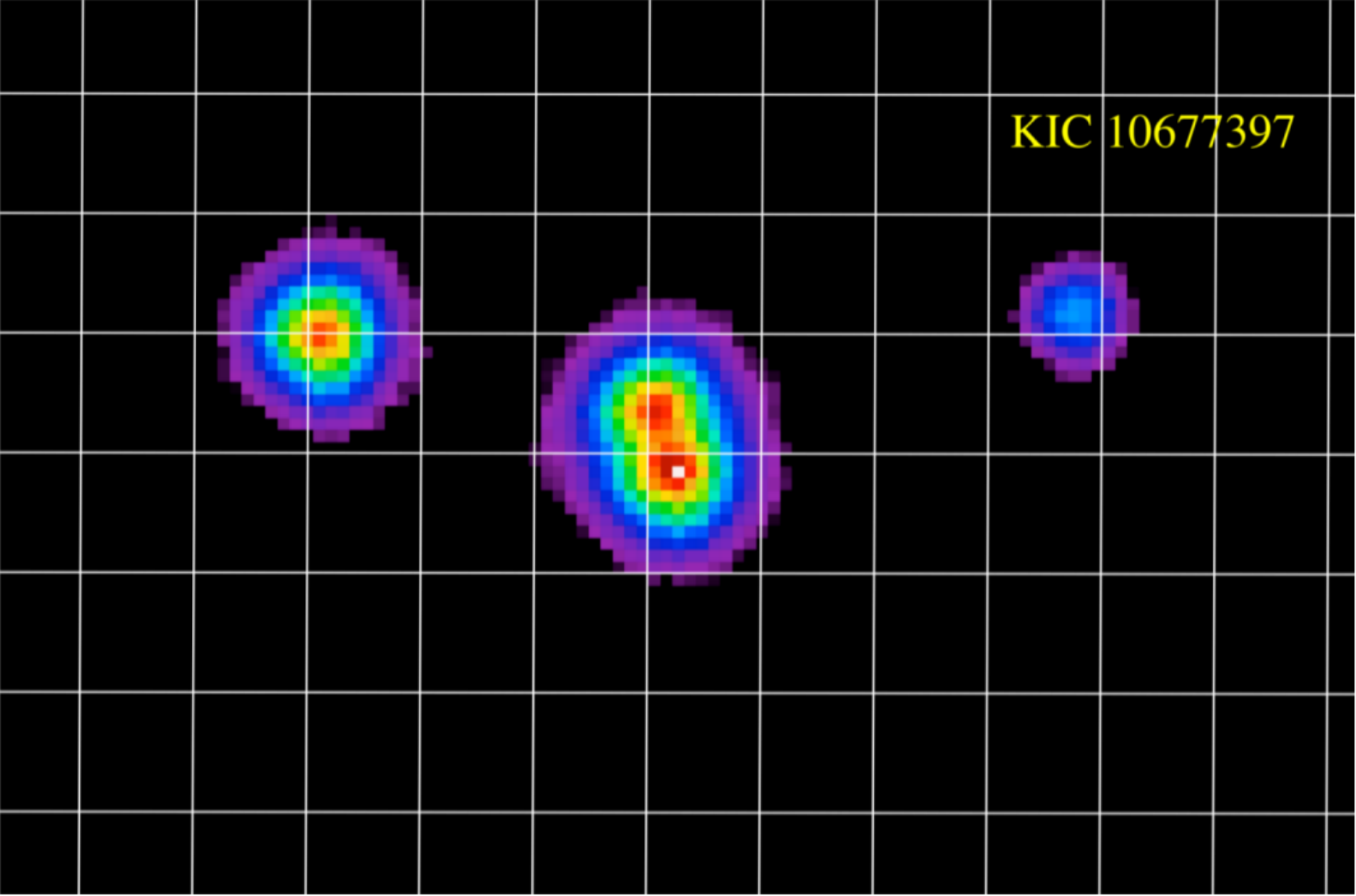} 
\caption{Selected set of UKIRT J-band images of {\em Kepler} M-stars which exhibit multiple rotation periods.  North is up and East is to the left.  From left to right and then down, they are KIC 4660255, KIC 7849619, KIC 9022001, KIC 9428095, KIC 10403228, and KIC 10677397. The white grid lines are separated by $2'' \times 2''$.  In all cases, North is up and East is to the left.  The images have been slightly smoothed with a 2D Gaussian with $\sigma = 0.2''$.  The color scale is related to the square root of the flux.  KIC 7849619, KIC 9428095, and KIC 10403228 appear as fairly close doubles, i.e., with separations of $\sim$$2.5''-3.0''$.  KIC 10677397 has a barely resolved stellar core (separation $\sim$$1''$) and has two possible stellar companions at 6$''$ and $7''$.  KIC 9022001 and KIC 4660255 are apparent doubles separated by $\sim$$4''$, and, in each case, the brighter star is elongated and may itself be a binary.}
\label{fig:UKIRT}
\end{center}
\end{figure*}

\section{Binary and Hierarchical M-Star Systems}
\label{sec:hierarchical}

We take the presence of two or more periodicities with incommensurate
periods, as described in the previous section, as evidence for the
presence of two or more rapidly rotating spotted stars within the
photometric aperture.  This conclusion seems inevitable, since these
cases are similar to the cases where only one periodicity is seen in a
given target, and since we know of no evidence, for any of these
periodicities, that indicates they are due to 
binary system modulations or stellar pulsations.

A simple statistical argument indicates that most of these targets
where two or more periods have been found must consist of multiple
physically related stars, i.e., stars in bound systems.  Of the
approximately 4000 M stars we studied, $\sim$200 or $\sim$$5$\% have at
least one rotation period shorter than 2 days, and about 100 or
$\sim$$2.5$\% have at least one rotation period shorter than 1 day.
Thus, given a {\em Kepler} target that exhibits a periodicity
with $P_{\rm rot} < 1$ day, the probability that a second periodicity
with $P_{\rm rot} < 2$ days will also be apparent is $\sim$$5$\% times
the chance that the image of a second M star is in the photometric 
aperture of the target.  The probability of finding two other M stars with 
$P_{\rm rot} < 2$ days by chance coincidence in the aperture must be 
lower than 0.3\%.  If we have examined 110 objects with $P_{\rm rot} < 1$ 
day, we should have found, by chance, {\em fewer} than 5 and 1/4 systems with 
either 2 or 3 M stars, respectively, in the same {\em Kepler} photometric 
aperture with $P_{\rm rot} < 2$ days.  The actual values are 30 and 3
(see Table \ref{tab:3Periods}), both
significantly in excess of what could be expected by chance.  Here we
have implicitly assumed that, on average, there is much less than one
serendipitously occurring second or third detectable M star in any given
photometric aperture.

In the above discussion and probability estimates for finding two or
more rapidly rotating stars within the same {\em Kepler} photometric
aperture, we made the assumption that the second and third stars were
also M stars.  If we loosen that assumption, then we can draw on the
broader statistics about rotation periods in the {\em Kepler} sample
found by Reinhold, Reiner, \& Basri (2013). They identified periodic
variations presumably due to spots on rotating stars, and thereby
compiled the rotation periods of all active {\em Kepler} stars.  Out
of their sample of 21,100 stars, they found 925 with periods $\lesssim                                                                 
2$ days.  Thus, rapidly rotating stars comprise only 4.5\% of stars
across the spectral types studied by {\em Kepler}.  This is
essentially the same value as we found for our M-star sample, and
thus the statistical argument presented above is again applicable,
i.e., there is a rather low probability of finding additional rapidly
rotating active stars with any of a range of spectral types in the
same aperture as the first M star with a short period.

We therefore adopt the working hypothesis that each M star target
having a detected periodicity with $P_{\rm rot} < 1$ day and also one
or more additional incommensurate rotational periodicities with
$P_{\rm rot} < 2$, must actually consist of multiple M stars bound in
a single system.

Finally, we note that 12 of the 37 systems listed in Table \ref{tab:3Periods} have two
short periods that are similar, i.e., they differ by no more than 4\% to 25\%.
There is one system where two periods differ by only 1\%.  This phenomenon 
can have two explanations.  The first is that the two periods originate in an M-star 
binary with two nearly equal mass stars of essentially the same age and spin-down 
history.  Thus, perhaps the closeness of a pair of rotation frequencies is not at all
unexpected.

The second possibility would be that two close periods come from
spots at different latitudes of a single star that is undergoing
differential rotation.  Reinhold et al.~(2013) have carried out an
extensive study of differential rotation in 40,660 active {\em Kepler}
stars. In some 18,600 of these stars they find two or more close
rotation periods which they take as evidence of differential
rotation and which they then use to derive limits on the differential rotation 
properties. Their Fig.~15 summarizes the measured horizontal shear
differential rotation parameter $\Delta \Omega$, which is defined as
the difference in rotation frequency between the equator and the pole.
For stars hotter than $\sim$6000 K, values of $\Delta \Omega$ are
often found to exceed 0.2 rad d$^{-1}$.  However, for cool stars, of
the type we are studying here, $\Delta \Omega$ averages about 0.07 rad
d$^{-1}$ and only rarely exceeds 0.1 rad d$^{-1}$.  We translate this
to a fractional differential rotation parameter, $\alpha$, for cool
stars of arbitrary rotation rate:
\begin{eqnarray}
\label{eqn:alpha}
\alpha \equiv \frac{\Delta \Omega}{\Omega_{\rm eq}} \simeq 0.01~ P({\rm days})~,
\end{eqnarray} 
where $\Omega_{\rm eq}$ is the equatorial rotation frequency of the
star, and $P=2\pi/\Omega_{\rm eq}$.  If this relation indeed holds
down to short periods (e.g., 1/2 day), then it implies that any two
short periods we detect which are different by more than a few percent
are {\em not} likely due to differential rotation.

In any event, the possibility that some of the periodicities we see
may arise on the {\em same} star should be kept in mind, though we do
not expect this to be the situation in many cases. Future tests with high 
resolution AO imaging and ground-based spectroscopy can be of further 
help for checking this possibility.

Reinhold et al.~(2013) also show (in their Fig.~8) that there is a strong trend for stars 
that are of spectral type F and earlier to rotate more rapidly than cooler stars which are 
presumably braked by magnetically constrained stellar winds (see, e.g., Mestel 1968; 
Skumanich 1972; Smith 1979; Zwaan 1981; Verbunt \& Zwaan 1981).  Such magnetic 
braking likely requires dynamo activity that is stronger in stars with convective envelopes.  
Therefore, according to Barnes \& Sofia (1996) and Barnes (2007), M stars that are rotating 
with periods shorter than 2 days are probably relatively young, i.e., $\lesssim$ one to a few 
hundred Myr (see also Sect.~\ref{sec:conclusions}). 

An alternative to the ``youth hypothesis'' is that the short periods
we see are indeed due to spots on rotating stars, but in tidally
locked close binaries.  One such example is V405~And with active
components consisting of M0V+M5V spectral types and rotational/orbital
periods of 0.465 day (Vida et al.~2009).  Since the light curves of our
short period systems do not exhibit obvious eclipses or ellipsoidal light
variations\footnote{In ellipsoidal variations the main peak in the FT occurs at twice the orbital 
frequency, $2\nu$, with very small contributions at $\nu$, and $3\nu$. 
We see no such cases. There is also the possibility that the amplitudes at $\nu$ and $3\nu$
are too low to be detected, leaving only the peak at $2\nu$.  Several cases of 
frequencies with no other harmonics are indeed seen, but these are clearly marked in Table 2.},
this would imply that we are viewing the systems at small
orbital inclination angles.  However, that would make the situation
even more extreme in the sense that each such short period would
itself require a binary.  Thus, in systems where we see three or four
short periods, this hypothesis would require 6-8 bound stars, which is
rather implausible.

\section{Imaging Evidence for Multiplicity}
\label{sec:images}

\subsection{UKIRT Images}

We have inspected the UKIRT J-band
images\footnote{\url{http://keplerscience.arc.nasa.gov/ToolsUKIRT.shtml}}
for the 37 systems listed in Table \ref{tab:3Periods} which exhibit two to four rotation periods.  
In general, for the UKIRT images of the {\em Kepler} field, one can easily
distinguish two stars of comparable brightness that are separated by
$\gtrsim 1.5''$, while for objects separated by less than $\approx 1''$, there 
is a single image.  Often, it is possible to discern that the image is elongated for 
stellar separations as small as $\approx 0.5''$.  Nineteen of the UKIRT J-band frames
appear to show single stellar images.  However, 11 show an apparent companion, or more 
than one companion, with a separation of $\lesssim 5''$, while 7 of these have separations 
$\lesssim 3''$.  Thus, there is tentative evidence for multiplicity in the UKIRT images of 
$\approx 30\%$ of the systems in Table \ref{tab:3Periods}.

In reviewing the evidence, it is good to keep in mind that a
separation of 1$\arcsec$ at typical distances of $\sim$200 pc
corresponds to a physical separation of 200 AU.  If, for sake of
argument, bound stellar pairs are taken to have orbital separations
between 0.01 AU and $10^5$ AU that are distributed uniformly in terms
of the logarithms of the separations (e.g., Dhital et al.~2010), then
fewer than half of them will have orbital separations $\gtrsim 200$
AU.  Thus, the UKIRT image multiplicity numbers noted above do not
rule out the possibility that all of
the 37 multiple M-star candidates are in bound hierarchical
systems (see e.g., Sect.~\ref{sec:AO} for some examples of closer
pairs).

\begin{figure}
\begin{center}
\includegraphics[width=0.97 \columnwidth]{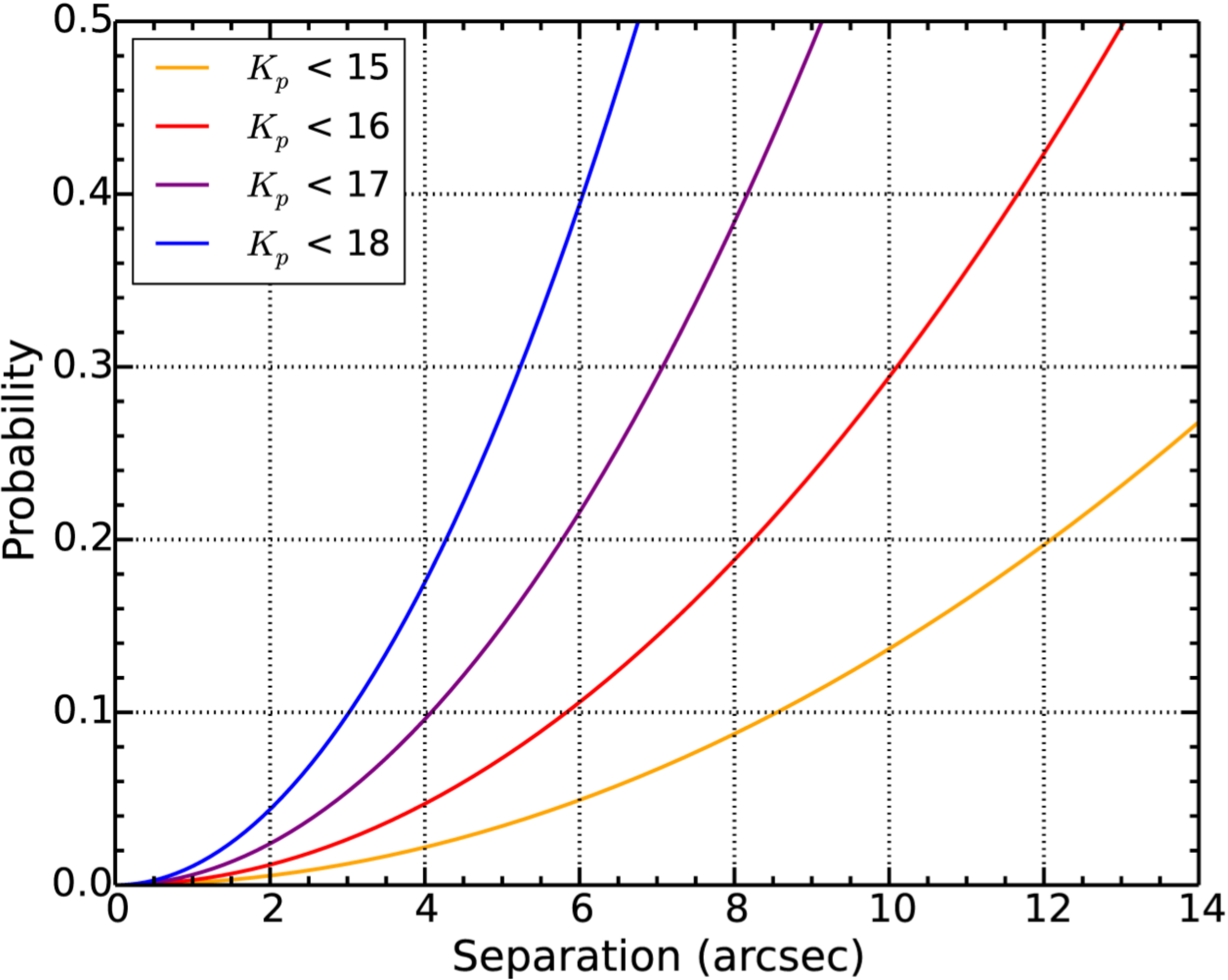} 
\caption{Probability of chance alignment of background or foreground stars with our 
{\em Kepler} M-star target, brighter than the indicated magnitude, as a function of separation. }
\label{fig:interlopers}
\end{center}
\end{figure}

\begin{figure*}
\begin{center}
\includegraphics[width=0.85 \textwidth]{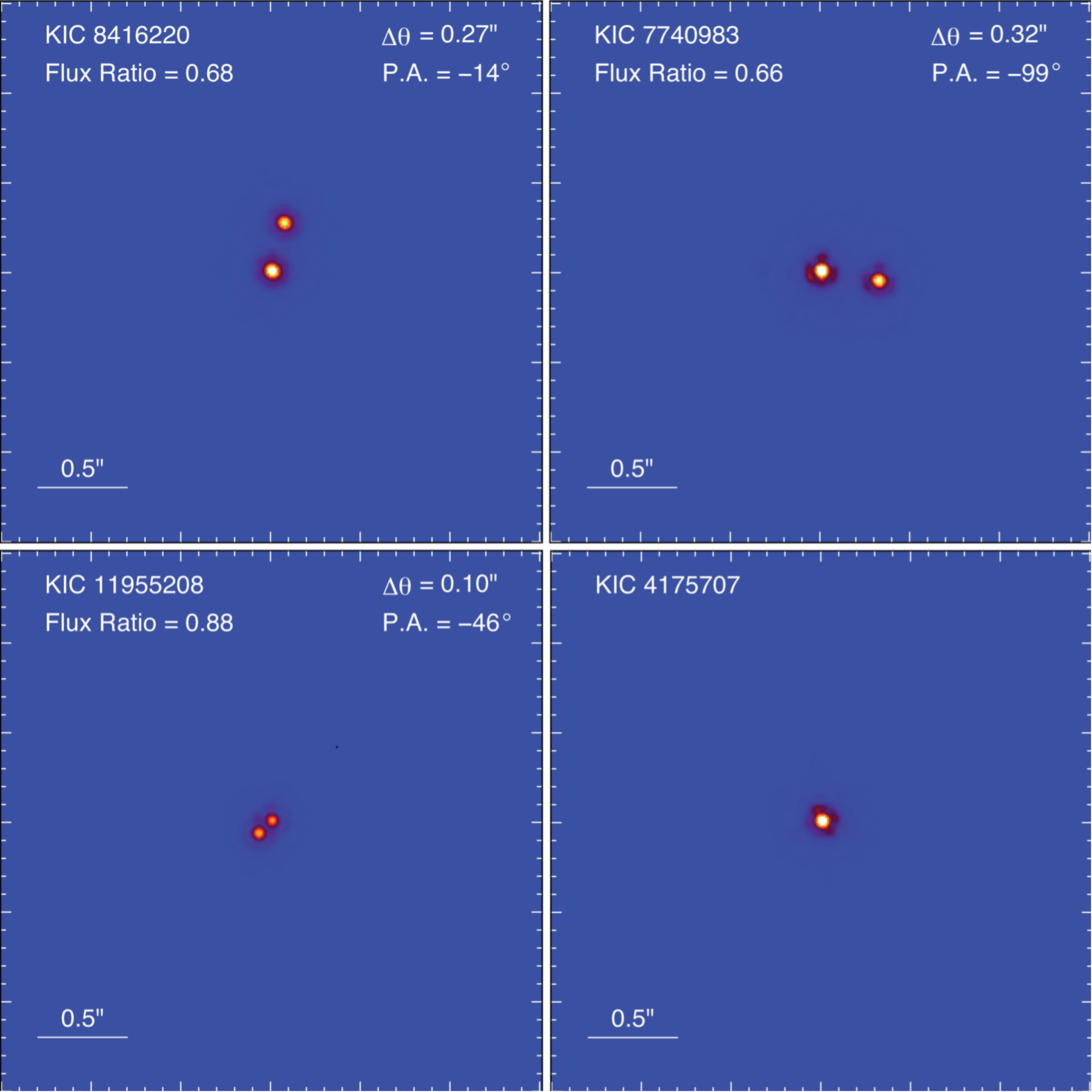} 
\caption{Keck adaptive optics images for four of our candidate multiple M-star systems.  North is up and East is to the left.  KIC 8416220, 11955208, 4175707 were recorded at $K^\prime$ band, while the image of 7740983 was taken in the $K_s$ band.  All except KIC 4175707 are obvious close twin binary M stars.  The angular separations between components ranges between 0.1$''$ and 0.32$''$, corresponding to physical separations of $\sim$ 20-65 AU at nominal distances of 200 pc. Flux ratios and position angles of the two components in degrees East of North are displayed where applicable.}
\label{fig:Keck}
\end{center}
\end{figure*}

In Fig.~\ref{fig:UKIRT} we present six examples of multiple M-star
candidates that show direct evidence for multiple stellar images
within the {\em Kepler} photometric aperture.  These systems are: KIC
4660255, 7849619, 9022001, 9428095, 10403228, and 10677397, and they
are included in Table \ref{tab:3Periods} along with their periods.
The {\em Kepler} magnitudes of these stars range from 15.3-16.1.  The
magnitudes and densities of stellar images in Fig.~\ref{fig:UKIRT}
indicate that most of these pairs of images are likely to be
physically related as opposed to chance alignments.  In
Fig.~\ref{fig:interlopers} we show the probability of interlopers
(i.e., background or foreground stars) coming within a given angular 
distance of any particular {\em Kepler} target star. 
These estimates were obtained by calculating a uniform stellar density
from the 10$^{\rm th}$ release of the {\em Kepler} Input Catalog
(KIC10)\footnote{\url{https://archive.stsci.edu/kepler/kic10/search.php}.}
for a given magnitude limit within a 115 square degree field of view   
(Borucki et al. 2010). The mean stellar densities are likely slight
overestimates since the KIC10 catalog includes roughly 36,000 
calibration stars outside the {\em Kepler} field of view. 
Although the stellar density is not uniform over {\em Kepler}'s field
of view---rather it varies with Galactic latitude---the distribution
of our multi-period sources are uniform in Galactic latitude
supporting our conclusion that the multi-period sources are most
likely bound companions rather than interlopers.    
We can see from Fig.~\ref{fig:interlopers} that for stars comparable
to, or up to 1-2 magnitudes fainter than, our $K_p$ = 15-16 magnitude
target M stars, it is worth considering stars out to $\sim$5$''$ as
possibly physically related to the M star in question. 

The white grid lines in Fig.~\ref{fig:UKIRT} are drawn with a spacing
of  $2\arcsec \times 2\arcsec$.  The images have been slightly
smoothed with a 2D Gaussian with $\sigma = 0.2''$.  KIC 7849619, KIC
9428095, and KIC 10403228 appear as fairly close doubles with
separations of $\sim$$2.5''-3.0''$.  KIC 10677397 has a barely
resolved stellar core (separation $\sim$$1''$) and has two possible
stellar companions at 6$''$ and $7''$.  KIC 9022001 and KIC 4660255
are apparent doubles separated by $\sim$$4''$, and, in each case, the
brighter star is elongated and may itself be a binary.  Finally, we
remind the reader that KIC 4660255 is the remarkable object that
exhibits four independent periods, all $< 1.2$ days.  This target is
discussed in detail in  Section~\ref{sec:4660}. 

Some of these separations seen in the UKIRT images
are sufficiently large, i.e., $2\arcsec - 5\arcsec$, that the point-spread-function (``PSF'')
fitting technique (Still \& Barclay 2013) may be useful for
identifying which of the stellar images is the source of a particular
periodicity.  This method is applied to KIC 4660255 in
Section~\ref{sec:4660}.

\subsection{Keck AO Images}
\label{sec:AO}
Images of four targets in our sample were obtained with the NIRC2
camera using the Keck II adaptive optics (AO) system during the past
ground-based observing season for the {\em Kepler} field. We observed
KIC 4175707, KIC 11955208, and KIC 8416220 on 2013 May 30 (UT) using
the laser guide star to close the AO system control loops (Wizinowich
et al.~2006). The targets were imaged in $K^\prime$ at low air mass
($\sec\,z < 1.1$) using a 3-point dither pattern with $2\arcsec$
dither spacings. Three images were obtained at each dither position
with integration times of 2, 3, and 2 s incorporating 10, 5, and 10
coadds per frame for a total integration time per target of 180, 135,
and 180 s, respectively. KIC 7740983 was imaged on the night of 2013
August 18 with Keck II/NIRC2 natural guide star AO (Wizinowich et
al.~2000) in the $K_s$ band. The same 3-point dither was performed for
this target with 3-s integrations, and 10 coadds per frame for a total
integration time of 180 s.

Each dither frame was corrected for bad pixels, sky subtracted, and
flat fielded. Centroids were obtained by fitting the core of the Airy
pattern with a two dimensional Gaussian function. The final images in
Figure~\ref{fig:Keck} show the medians of the aligned dither stacks for
each target.

Each of these four targets shows a pair of rotation periods (in days)
that are shorter than 1 day, viz., \{0.57,0.72\}, \{0.40,0.52\},  \{0.56,0.70\}, 
and \{0.34,0.42\}, for KIC 8416220, KIC 7740983, KIC 11955208, and 
KIC 4175707, respectively.  Amazingly, three of the four systems show 
close twin pairs of M stars in what are almost certainly bound binaries if 
not higher-order multiples. Only KIC 4175707 still appears single at the
$0.05^{\prime\prime}$ level.

\begin{figure}
\begin{center}
\includegraphics[width=0.97 \columnwidth]{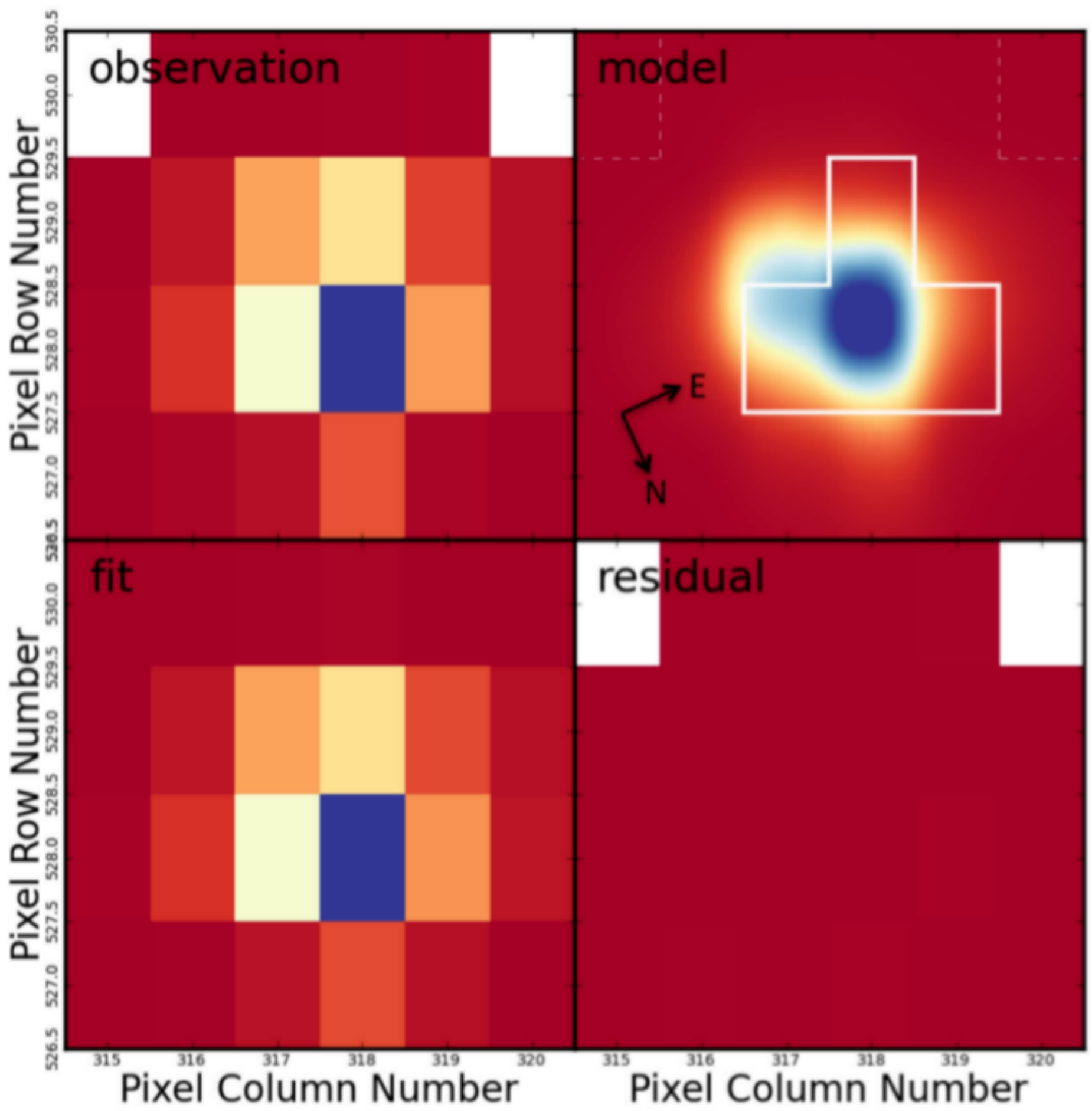} 
\caption{Typical 29.4-min.\,long-cadence observation of KIC 4660255, at the pixel level (each pixel is $4'' \times 4''$), for one particular frame of $\sim$50,000 exposures. The pixel mask for KIC 4660255 contains at least two stars -- the bright blue pixel at \{318, 528\} as well as a fainter companion star that is about 4$''$ to the left (see Fig.~\ref{fig:UKIRT}).  The best PSF model fit is plotted in the top-right, while that fit, rebinned across the detector pixels, is compared at the lower-left. The lower-right panel contains the residuals between the data and the best fit.  All images are plotted on a linear intensity scale.  PSF-derived light curves are constructed by repeating this fit for every exposure over quarters 1 to 16.  The objective is to disentangle the fluxes between the two stars within the pixel mask. The approximate orientation of the pixels with respect to the sky is indicated by the compass in the upper right panel.  The inverted ``T''-shaped region in the same panel indicates a typical pixel-level photometric aperture for this target used for conventional analyses.}
\label{fig:PSF}
\end{center}
\end{figure}

\begin{figure}
\begin{center}
\includegraphics[width=0.97 \columnwidth]{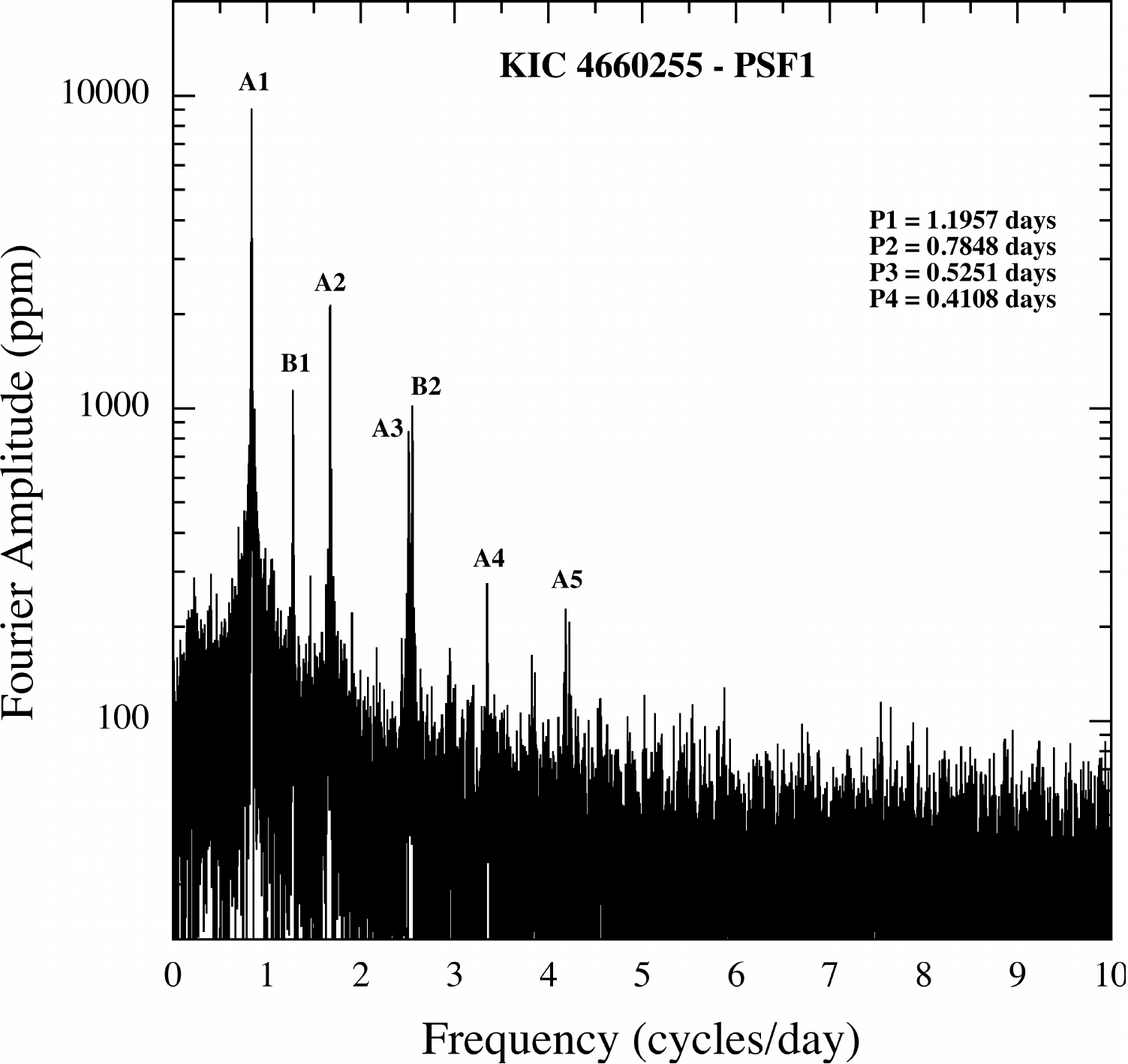} \vglue0.2cm 
\includegraphics[width=0.97 \columnwidth]{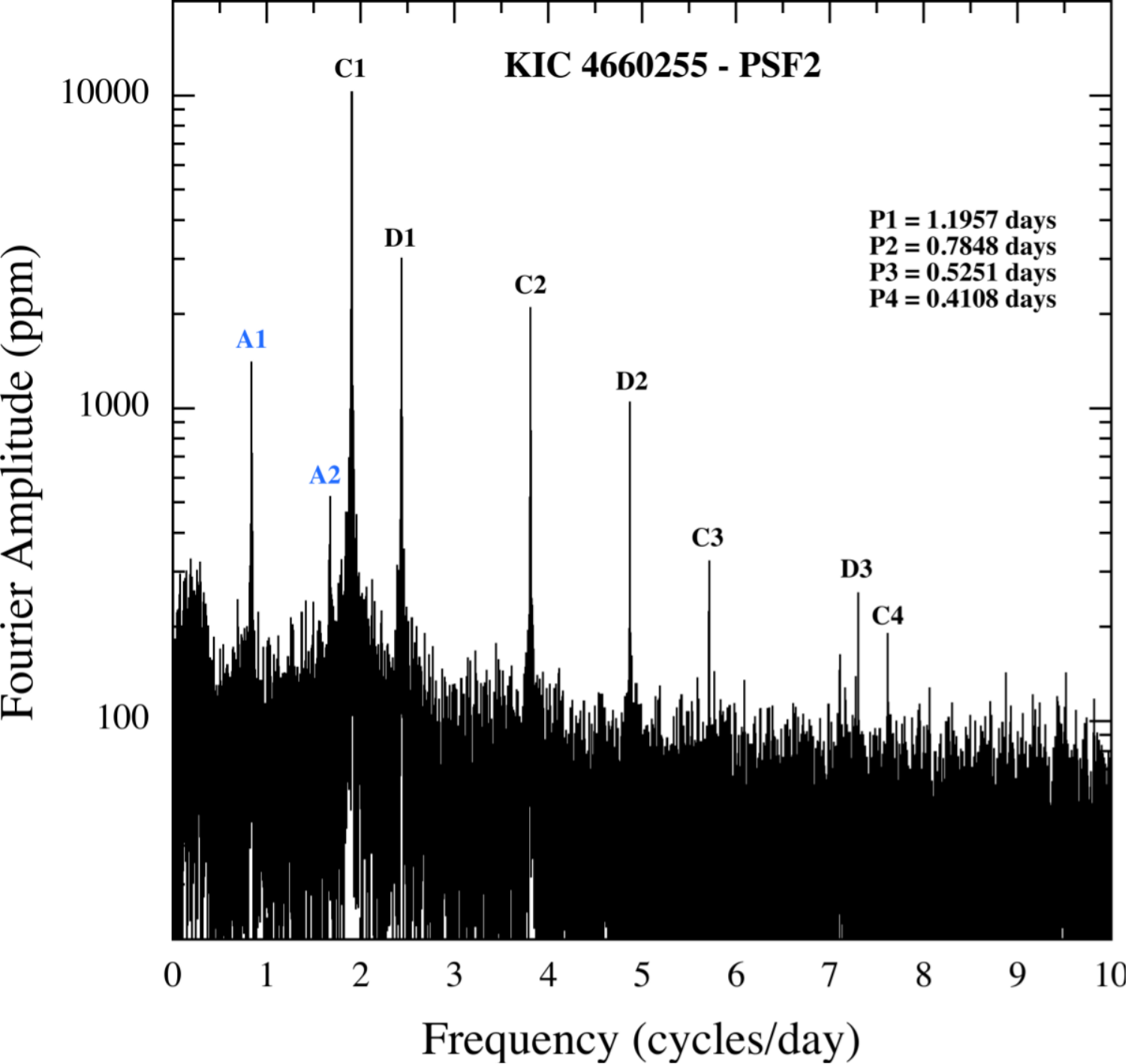} 
\caption{Fourier transforms of the PSF analyzed data trains for KIC 4660255.  Top: This is the PSF-extracted data train for the brighter of the two stars within the pixel mask.  Note that of the four periodicities in this {\em Kepler} target, only the longer two periods show up clearly.  Bottom: PSF-extracted data train for the fainter of the two stars within the pixel mask.  Here, the shorter two periods show up prominently, with only a small ``leakage'' of the brighter star and its period of 1.19 days.  Thus, we associate two rotation periods with the brighter star and the other two with the fainter star.}
\label{fig:FT466}
\end{center}
\end{figure}

\section{Special Case of KIC 4660255}
\label{sec:4660}

In this section we examine KIC 4660255, which exhibits four rotational
periodicities all with periods shorter than 1.2 days.  As the upper
left panel of Fig.~\ref{fig:UKIRT} shows, at the location of KIC
4660255 there are two point-like images that are separated by 4.2$''$,
and, hence, are well within the photometric aperture.  Still \& Barclay
(2014) have devised a point-spread function (``PSF'') fitting technique
that can utilize data at the pixel level to compute light curves for each
of several stars within the photometric aperture.  In brief, we assume here that the
two stars can be characterized as two point-spread-functions with two unknown 
fluxes and shifts of the photometric aperture relative to the sky, $\Delta \alpha$ 
and $\Delta \delta$ (see Fig.~\ref{fig:PSF}). We then minimize upon the uncertainty-weighted 
residuals between the pixel-level data and the fit in order to yield the flux and pixel position of 
both stars for every one of 48,100 exposures across quarters Q1-16. This technique was 
demonstrated in the case of KOI 2700 to determine which of two stars within the photometric 
aperture was the source of planet transits (Rappaport et al.~2013).

We have carried out a PSF analysis for KIC 4660255 to try to identify from which 
of the two well-separated images (see Fig.~\ref{fig:UKIRT}) the various rotation 
frequencies arise.  We show in Fig.~\ref{fig:PSF} an example of a single frame of 
pixel-level data for this object.  The upper left panel shows the $4'' \times 4''$ {\em Kepler} 
pixels within the pixel mask for KIC 4660255 for a single frame of the 50,575 available 
for this object.  The best fitting PSF model, comprised of two stars, is shown in the upper 
right panel.  When this model is integrated over the pixels, the ``fit'' is shown in the lower 
left panel.  The residuals with respect to the data from that frame, are shown in the 
lower-right panel.  This process was repeated for each of the 50,575 frames for KIC 4660255, 
thereby yielding two photometric time series, one for each star.  

The resultant FTs for the time series extracted from the two stars
within the aperture of KIC 4660255 are shown in Fig.~\ref{fig:FT466}.
The top panel shows the FT for the brighter of the two stars (the more
northerly one).  Only the $A$ and $B$ frequencies are present at a
detectable level.  The corresponding periods are 1.1958 days and
0.7859 days.  No evidence is seen for the shorter $C$ and $D$ periods.
In contrast, the bottom panel in Fig.~\ref{fig:FT466} shows the $C$
and $D$ periods, 0.5251 and 0.4108 days, very clearly. In addition,
the first two harmonics of the $A$ periodicity leak through weakly --
at about 1/5 of the amplitude that they exhibit in the upper panel.
Thus, the PSF fitting technique appears to have cleanly distinguished
the four periodicities as being from two different stellar sources.
In turn, we can marginally discern from Fig.~\ref{fig:UKIRT} that the
northerly image which is host to periods $A$ and $B$ likely comprises
two images, and we propose that each of those is the source
of one of the $A$ and $B$ periods.  By analogy, we hypothesize that
the fainter, more southerly image is itself a binary, each star of
which is rotating with the $C$ or $D$ period.

\section{Summary and Conclusions}
\label{sec:conclusions}

In this work we have searched the {\em Kepler} photometric data base of M stars for short 
rotation periods.  We found 178 {\em Kepler} targets with 211 different periods of $P_{\rm rot} < 2$ 
days and 110 systems with 127 different periods with $P_{\rm rot} < 1$ day.  Some 30 of these 
objects exhibit two or more rotation periods where at least one has $P_{\rm rot} < 1$ day and 
the other has $P_{\rm rot} < 2$ days.  We also find several systems with three or more rotation 
periods where at least one has $P_{\rm rot} < 1$ day and the others have $P_{\rm rot} < 2$ day.  
There are a sufficient number of these with more than one short rotation period to allow us to 
argue that they are likely young, physically related, binary or hierarchical triple systems.   

At least 6, and perhaps as many as 14, of these M-star systems show,
in the UKIRT {\em Kepler} region survey J-band images, multiple
stellar images\footnote{These 14 systems are ones with another stellar image
within $\sim$$5''$ of the target and within $\sim$$| \Delta J | \lesssim 2$.} that may represent physically related stars.  
Three more systems were clearly resolved in Keck AO images of four systems.
These are almost certainly physically bound twin M-star binaries.  It
seems quite likely that additional AO imaging in the future will reveal that
more of these 30 multiple-period systems contain multiple stellar images.  Spectroscopy could
further reveal closer binary members.  

If most of these M-star systems with multiple short periods indeed turn out to be multiple bound 
M-stars, this could prove a valuable way of discovering young hierarchical M-star systems.  We 
suggest that this approach may also be applicable to K and G stars.

In the process of conducting this study we found that approximately 5\% of all M stars are rotating 
with periods shorter than 2 days.  If we utilize standard models of contraction onto the main 
sequence, with the consequent {\em spinup} of the star due to conservation of angular momentum, 
and the loss of systemic angular momentum due to magnetic braking (see, e.g., Kawaler 1988; 
Chaboyer et al.~1995; Barnes \& Sofia 1996; Irwin et al.~2011) we can check whether M stars 
spend sufficient time rotating rapidly to allow for 5\% of them to have $P_{\rm rot} \lesssim 2$ days.  
These models typically take the magnetic braking torque to be proportional to $\omega_{\rm rot}^3$ 
when $\omega_{\rm rot} < \omega_{\rm sat}$, where $\omega_{\rm  sat}$ is a ``saturation'' frequency, 
above which the torque is proportional to $\omega_{\rm rot} \omega_{\rm sat}^2$.  As an example of 
a model spin history of an M star, see Fig.~13 of Irwin et al.~(2011; dashed curve).  We see that a 
minimum rotation period of 0.3 days is attained at $\sim$10 Myr, and the period lengthens to more 
than 1 day at 300 Myr and 2 days at 400 Myr.  Thus, at least for this specific example, which involves 
numerous uncertainties, the fraction of time that a $10^{10}$ year old M star has spent with 
$P_{\rm rot} \lesssim 2$ days is about 4\% of its lifetime.  This is reassuringly consistent with the 
fraction of M stars that we find are rotating rapidly.

\acknowledgements We acknowledge several useful discussions about the particular M star(s) 
KIC 7740983 with Robert Szabo and Katrien Kolenberg.  We also thank Cristina Rodriguez-Lopez, 
Jim MacDonald, J\'er\^ome Quintin, Alex Brown, G\"unter Houdek, and Lucianne Walkowicz for 
important discussions about the possibility of observing M-star pulsations with {\em Kepler}. 
Arthur Delarue wrote a very helpful code for automatically extracting significant incommensurate 
frequencies from Fourier transforms.  We thank Sasha Hinkley and Benjamin Montet for 
performing a subset of our Keck AO observations. We are grateful to the {\it Kepler} team for 
providing such valuable data to the community. D.H.~acknowledges NASA support through the 
Kepler Participating Scientist Program under Grant NNX14AB92G.  R.S.O.~acknowledges 
support through the Kepler Participating Scientist Program and the NASA Origins Program under 
Grants NNX12AC76G and NNX11AG85G.  P.S.M.~acknowledges support from the Hubble Fellowship 
Program, provided by NASA through Hubble Fellowship grant HST-HF-51326.01-A awarded by the 
STScI, which is operated by the AURA, Inc., for NASA, under contract NAS 5-26555.  
G.H.~is grateful for support by the Polish NCN 
grant 2011/01/B/ST9/05448.  K.O.~and K.V.~acknowledge support from the Hungarian OTKA grants 
K-81421 and K-109276, and from ``Lend\"ulet-2012'' Young Researchers' Programs of the 
Hungarian Academy of Sciences.  This research has made use of the NASA Exoplanet Archive, 
and the Mikulski Archive for Space Telescopes (MAST).  We made use of J-band images that 
were obtained with the United Kingdom Infrared Telescope (UKIRT) which is operated by the 
Joint Astronomy Centre on behalf of the Science and Technology Facilities Council of the U.K. 
Some of the data presented herein were obtained at the W.M. Keck Observatory, which is 
operated as a scientific partnership among the California Institute of Technology, the University 
of California and the National Aeronautics and Space Administration. The Observatory was made 
possible by the generous financial support of the W.M. Keck Foundation. We acknowledge the 
very significant cultural role and reverence that the summit of Mauna Kea has always had within
 the indigenous Hawaiian community, and we are most fortunate to have the opportunity to 
 conduct observations from this mountain.

\newpage

\begin{center}
\input{Table1.tbl}
\end{center}

\begin{center}
\input{Table2.tbl}
\end{center}

\end{document}

%% file: Table1.tbl.tex
\begin{deluxetable}{cc|cc|cc|cc}
\tablewidth{0pt}
\tabletypesize{\tiny}
\tablecaption{\label{tab:1Period} {\em Kepler} M Stars Exhibiting a Short Rotation Period
}
\tablehead{
	\colhead{KIC} &
	\colhead{Period} &
	\colhead{KIC}  & 
	\colhead{Period} &
	\colhead{KIC} &
	\colhead{Period} &
	\colhead{KIC} &
	\colhead{Period}  \\
}
\startdata 
 1572802& 0.3711&  6592335& 0.4119&  8565914& 1.4183& 10790812& 1.0629\\ 
 2300039& 1.7083&  6664639& 1.4482&  8611876& 1.5797& 10790838& 0.9173\\ 
 2449101& 0.7399&  6715960& 0.9185&  8672278& 1.8549& 10796551& 1.1485\\ 
 2557669& 1.8595&  6752578& 0.2601&  8673358& 1.0292& 10803430& 1.5207\\ 
 3130391& 1.2301&  6762389& 0.3855&  8681527& 0.5775& 10975238& 1.9487\\ 
 3439791& 0.4321&  6928206& 1.2839&  8873575& 0.6702& 11031746& 1.2109\\ 
 3454793& 0.6682&  6936046& 0.5274&  8909833& 1.2812& 11042875& 1.5658\\ 
 3642335& 0.6725&  6949412& 1.6453&  8935942& 1.6582& 11091336& 1.1947\\ 
 3732401& 1.7990&  7110077& 0.9717&  9022001& 0.6989& 11124203& 0.9596\\ 
 3748172& 1.0516&  7269729& 0.3346&  9041966& 0.1477& 11140425& 1.1879\\ 
 3757251& 0.2158&  7431659& 1.0507&  9075708& 0.5131& 11147271& 1.8794\\ 
 3831911& 0.5621&  7434110& 0.4014&  9083354& 1.1341& 11189348& 0.5060\\ 
 3935499& 0.9214&  7445605& 1.0782&  9091897& 0.2399& 11305240& 0.7808\\ 
 3962433& 0.4025&  7448057& 0.3873&  9142641& 1.3747& 11343461& 1.6167\\ 
 4036313& 1.5816&  7449695& 0.5610&  9142714& 1.1285& 11349556& 1.6941\\ 
 4077867& 0.4777&  7547969& 0.6802&  9205855& 0.9144& 11390683& 1.1295\\ 
 4246255& 1.3177&  7592990& 0.4426&  9268481& 0.5802& 11446073& 1.0393\\ 
 4264634& 0.7541&  7678417& 0.6267&  9335198& 0.2898& 11447564& 0.3812\\ 
 4473355& 0.2060&  7686474& 1.4843&  9395840& 1.0973& 11498106& 0.5708\\ 
 4545041& 0.3326&  7733540& 1.5488&  9479539& 1.8576& 11521274& 0.6653\\ 
 4660255& 0.4109&  7740983& 0.4036&  9519275& 0.5797& 11605209& 1.8622\\ 
 4951466& 0.1349&  7741987& 1.2612&  9590249& 0.2642& 11668095& 0.9156\\ 
 5083330& 1.9550&  7743830& 0.3749&  9702550& 1.6270& 11702167& 0.3798\\ 
 5095098& 0.7414&  7800087& 0.4722&  9705079& 0.5377& 11722217& 0.1643\\ 
 5182822& 0.3391&  7847566& 0.3806&  9761113& 0.6806& 11760021& 0.7565\\ 
 5184487& 1.1051&  7849619& 0.3092&  9784820& 1.1005& 11855334& 0.8173\\ 
 5341666& 0.3631&  7973675& 0.4572&  9899900& 0.4145& 11855853& 1.4325\\ 
 5360129& 1.3519&  8012943& 0.5726&  9933464& 1.2234& 11876220& 1.4600\\ 
 5435958& 0.8370&  8057610& 1.7332&  9992083& 0.4197& 11955208& 0.5637\\ 
 5621528& 0.7991&  8075991& 1.2932& 10027247& 0.5887& 12022407& 0.7976\\ 
 5685704& 0.5129&  8098178& 0.9874& 10324374& 1.7311& 12058533& 0.3442\\ 
 5771150& 0.5049&  8098228& 0.5472& 10384891& 0.4079& 12060710& 0.3788\\ 
 5937264& 0.6037&  8107903& 1.5406& 10403228& 0.2369& 12105694& 1.9567\\ 
 5938531& 0.8800&  8150479& 0.3805& 10412044& 1.4002& 12105867& 1.1736\\ 
 5951140& 1.0551&  8183594& 0.2980& 10462462& 0.8696& 12203082& 0.5263\\ 
 5952378& 0.4528&  8248415& 0.9892& 10469305& 0.8971& 12207432& 1.6982\\ 
 5962956& 1.3998&  8257134& 0.2995& 10515986& 0.7472& 12258225& 0.9423\\ 
 6102091& 0.5162&  8325962& 0.5721& 10518758& 1.3394& 12304013& 0.3498\\ 
 6117832& 0.6047&  8414250& 0.7446& 10536761& 1.0255& 12356535& 0.8096\\ 
 6122790& 1.1193&  8415004& 1.2297& 10552016& 1.5089& 12365719& 0.8521\\ 
 6183736& 1.1819&  8416220& 0.5660& 10553513& 0.2547& 12505054& 0.4397\\ 
 6370174& 0.8606&  8417053& 1.4751& 10584063& 1.4280& 12784183& 0.2087\\ 
 6425928& 0.3255&  8447096& 1.2361& 10587237& 0.9603& 12835232& 1.8951\\ 
 6469920& 0.9717&  8454353& 1.4905& 10677397& 0.3115&        & \\ 
 6529445& 0.2231&  8474897& 0.9954& 10710753& 0.6493&        & \\ 
\enddata
\tablecomments{178 {\em Kepler} targets exhibiting at least one starspot rotation period shorter than 2 days.  If more than one period is present, only the shortest period is listed here.  Periods are in days.  For systems with more than one short rotation period see Table \ref{tab:3Periods}}
\end{deluxetable}

%% file: Table2.tbl.tex
\begin{deluxetable}{cccccccccc}
\tablewidth{0pt}
\tabletypesize{\tiny}
\tablecaption{\label{tab:3Periods} {\em Kepler} M Stars Exhibiting Two or More Short Rotation Periods
}
\tablehead{
	\colhead{Object} &
	\colhead{$\alpha_{J2000}$} &
	\colhead{$\delta_{J2000}$}  & 
	\colhead{$K_p$} &
	\colhead{$T_{\rm eff}$} &
	\colhead{$P_A$} &
	\colhead{$P_B$} &
	\colhead{$P_C$} &
    	\colhead{$P_D$} &
	\colhead{Imaging} \\
	(1)  & (2) & (3) & (4) & (5) & (6) & (7) & (8)  & (9) & (10)
}
\startdata 
3454793 & 19h 36m 49.43s  & 38d 31m 47.50s & 15.5 & 3431 & 0.6681 & 0.9673 & ... & ... & UKIRT: elongated (?)\\ 
3757251 & 19h 36m 14.20s  & 38d 50m 19.25s & 15.8 & 3623 & 0.2158 & 0.4485 & 15.40 & ... & UKIRT: 2.6$''$ \& 4.2$''$ \\ 
3831911 & 19h 00m 13.23s  & 38d 59m 06.04s & 15.0 & 3953 & 0.5670 & 1.8343 & 30.484$^\dag$ & ...  & UKIRT: 4.0$''$ \\ 
3962433 & 19h 34m 28.26s  & 39d 01m 38.78s & 15.8 & 3741 & 0.4024 & 0.6378 & ... & ... & UKIRT: single \\ 
4077867 & 19h 45m 41.35s & 39d 06m 34.56s & 15.8 & 3490 & 0.4776 & 0.9798 & 19.35 & ... & UKIRT: 4.0$''$ \\ 
4175707 & 19h 42m 47.72s  & 39d 16m 00.66s   & 15.0 & 3825 &  0.3406 & 0.4163 & ... & ... & Keck AO: single \\ 
4264634 & 19h 28m 34.15s   & 39d 18m 00.76s & 14.8 & 3939 & 0.7536 & 0.8865 & 12.38 & ... & UKIRT: single \\

4545041 &  19h 04m 19.82s & 39d 37m 18.16s & 15.9 & 3329 & 0.3321 & 0.4045 & ... & ... &  UKIRT: single \\ 
4660255 & 19h 33m 17.12s  & 39d 42m 32.62s  & 15.4 & 3917 & 0.4108 & 0.5251 & 0.7859 & 1.1958 &  UKIRT: elongated \& 4.2$''$ \\ 
5182822 & 19h 22m 00.25s  & 40d 22m 13.08s  & 15.7 & 3900 & 0.3390 & 1.1660 & 15.51 & ... & UKIRT: single \\ 
6425928 & 19h 01m 05.40s & 41d 51m 29.38s & 14.5 & 3862 & 0.3254 & 0.3942 & 2.4018 & 2.9842 &  UKIRT: single \\
6529445 & 19h 36m 05.46s  & 41d 56m 23.42s  & 15.7 & 3746 & 0.2231 & 0.2362 & ... & ... & UKIRT: elongated \& 4.7$''$\\  

6936046 & 19h 10m 16.10s  & 42d 26m 36.49s & 15.0 & 3772 & 0.5273 & 1.2327 & 13.21 & ... & UKIRT: $6.6''^\ddagger$ \\
7110077 & 19h 17m 35.66s  & 42d 41m 37.28s & 15.6 & 3598 & 0.9716 & 1.0390 & 22.08 & ... & UKIRT: 2 faint compan. at $\sim$1.3$''$ \\ 
7434110 &  19h 13m 24.06s & 43d 02m 47.00s  & 16.0 & 3311 & 0.4014 & 1.8070 & ... & ... & UKIRT: single \\
7448057 & 19h 31m 04.12s  & 43d 05m 54.96s & 15.6 & 3500 & 0.3872* & 1.9594 & ... & ... & UKIRT: single \\ 
7449695 & 19h 32m 58.53s  & 43d 04m 42.92s  & 16.0 & 3787 & 0.5609 & 1.6635 & ... & ... & UKIRT: single \\

7740983 & 19h 07m 57.37s & 43d 29m 56.08s  & 14.7 & 3727 & 0.4036 & 0.5190 & 3.33 & ... & Keck AO: 0.32$''$ binary \\
7849619 & 19h 58m 02.71s  & 43d 34m 02.06s & 15.5 & 3799 & 0.3092 & 0.3298 & 12.91 & ... &  UKIRT: 2.6$''$ \\ 
7973675 & 19h 45m 11.65s  & 43d 45m 26.50s  & 15.1 & 3506 & 0.4572 & 0.7764 & 17.40 & ... & UKIRT: single \\ 
8325962 &  19h 58m 46.21s  & 44d 15m 05.62s & 15.0 & 3946 & 0.1690$^*$ & 0.5719 & 7.721& 12.34 & UKIRT: 6.3$''$\\ 
8416220 & 19h 00m 57.66s  & 44d 28m 27.91s & 15.1 & 3227 & 0.5659 & 0.7151 & ... & ... & Keck AO: 0.27$''$ binary  \\

8701179 & 19h 44m 31.15s  & 44d 53m 48.08s & 15.9 & 3900 & 0.7046 & 0.7874 & 16.18 & ... & UKIRT: single \\
8873575 & 19h 07m 14.00s  & 45d 07m 40.12s  & 16.5 & 3906 & 0.6737 & 1.0226 & 11.88 & ... &  UKIRT: single \\
8909833 & 19h 57m 23.98s & 45d 09m 38.92s  & 15.6 & 3903 & 1.2802 & 7.385 & 11.80 & ... & UKIRT: $7'' \& 11''^\ddagger$ \\
9022001 & 19h 27m 18.51s  & 45d 22m 42.02s  & 16.0 & 3648 & 0.6989$^{**}$ & 0.7055$^{**}$ & 14.48 & ... & UKIRT: elongated \& $4.0''$ \\
9268481 & 19h 02m 03.70s  & 45d 44m 06.76s  & 15.6 & 3198 & 0.5802$^*$ & 1.5780 & 1.8638 & ... & UKIRT: single \\

9428095 & 19h 59m 37.83s & 45d 56m 04.78s  & 15.9 & 3431 & 0.6289 & 0.6974 & 23.04 & ... & UKIRT: 3.0$''$ \\ 
9519275 & 19h 14m 06.86s & 46d 07m 18.77s &15.5 & 3578 & 0.5798 & 1.8706 & 17.71 & ... & UKIRT: 6$''$ \\ 
9590249 & 19h 30m 58.54s  & 46d 13m 12.76s  & 15.8 & 3855 & 0.2640$^*$ & 2.1665 & 12.0 & ... & UKIRT: single \\
10403228 & 19h 24m 54.41s   & 47d 32m 59.93s  & 16.1 & 3386 & 0.2367 & 0.2461 & 34.6 & ... & UKIRT: 2.8$''$ \\ 
10584063 & 18h 57m 06.17s  & 47d 49m 28.67s  & 15.7 & 3316 & 1.397$^*$ & 1.5988 & 18.72 & ... & UKIRT: single \\ 
10677397 & 19h 45m 09.30s & 47d 54m 59.15s  & 15.3 & 3387 & 0.3115* & 0.4743 & 7.803  & ... & UKIRT: double \& 6$''$ \& 7$''$\\ 

11955208 & 19h 06m 43.50s & 50d 20m 08.77s  & 15.7 & 3604 & 0.5637 & 0.7028 & 1.1855 & ... & Keck AO: 0.1$''$ binary  \\
12203082 & 19h 12m 51.12s & 50d 48m 00.61s  & 15.3 & 3804 & 0.5263$^*$ & 0.6592 & 11.75 & ... & UKIRT: single \\
12258225 & 19h 25m 41.94s  & 50d 56m 41.82s & 14.7 & 3500 & 0.9416 & 1.0141 & ... & ... & UKIRT: single \\
12304013 & 19h 17m 40.27s  & 51d 00m 16.92s  & 15.2 & 3465 & 0.3497 & 2.3491 & 19.12 & ... & UKIRT: single \\
\enddata
\tablecomments{* = inferred from only the base frequency and no higher harmonics; ** = sufficiently close that these may result from differential rotation of a single star; $\dag$ - the period of an eccentric binary component of the system; (1) KIC ID, (2) Right Ascension, (3) Declination, (4) {\em Kepler} magnitude, $K_p$, (5) composite $T_{\rm eff}$, (6)--(9) rotation period in days, (10) Comments on the stellar neighbors inferred from the UKIRT J-band images, as well as the Keck AO images -- where available; angular distances to some of the neighboring stars are noted. $\ddagger$ - the other stars are of a distinctly different, i.e., hotter, spectral type.}
\end{deluxetable}